\documentclass[aps,prl,twocolumn,amsmath,amssymb,floatfix,superscriptaddress]{revtex4-1}
\usepackage{tabularx}
\usepackage{bm}
\usepackage{euscript}
\usepackage{epsfig,psfrag,subfigure}
\usepackage{graphicx}
\usepackage{color}
\usepackage{amsfonts}
\usepackage{exscale}
\usepackage{amsbsy}
\usepackage{multirow}
\usepackage{hyperref}
\hypersetup{colorlinks=false,linkcolor=black}
\usepackage[]{lineno}
\usepackage{comment}
\usepackage{ulem}
\usepackage[dvipsnames]{xcolor}
\def\be{\begin{equation}}       \def\ee{\end{equation}}
\def\bea{\begin{eqnarray}}      \def\eea{\end{eqnarray}}

\newcommand{\PreserveBackslash}[1]{\let\temp=\\#1\let\\=\temp}
\newcolumntype{C}[1]{>{\PreserveBackslash\centering}p{#1}}

\begin{document}
\title{Gapless spin liquid and pair density wave of the Hubbard model on three-leg triangular cylinders}

\author{Cheng Peng}
\affiliation{Stanford Institute for Materials and Energy Sciences, SLAC National Accelerator Laboratory and Stanford University, Menlo Park, California 94025, USA}

\author{Yi-Fan Jiang}
\affiliation{Stanford Institute for Materials and Energy Sciences, SLAC National Accelerator Laboratory and Stanford University, Menlo Park, California 94025, USA}

\author{Yao Wang}
\affiliation{Department of Physics and Astronomy, Clemson University, Clemson, SC 29631, USA}

\author{Hong-Chen Jiang}
\email{hcjiang@stanford.edu}
\affiliation{Stanford Institute for Materials and Energy Sciences, SLAC National Accelerator Laboratory and Stanford University, Menlo Park, California 94025, USA}

\begin{abstract}
We study the ground state properties of the Hubbard model on three-leg triangular cylinders using large-scale density-matrix renormalization group simulations. At half filling, we identify an intermediate gapless spin liquid phase between a metallic phase at weak coupling and Mott insulating dimer phase at strong interaction, which has one gapless spin mode and algebraic spin-spin correlations but exponential decay scalar chiral-chiral correlations. Upon light doping the gapless spin liquid, the system exhibits power-law charge-density-wave (CDW) correlations but short-range single-particle, spin-spin, and chiral-chiral correlations. Similar to CDW correlations, the superconducting correlations are also quasi-long-ranged but oscillate in sign as a function of distance, which is consistent with the striped pair-density wave. When further doping the gapless spin liquid phase  or doping the dimer order phase, another phase takes over, which has similar CDW correlations but all other correlations decay exponentially.
\end{abstract}

\maketitle

Pair density wave (PDW) is a superconducting (SC) state in which Cooper pairs have finite momentum and the order parameter varies periodically in space in such a way that its spatial average vanishes.\cite{LeeSS2005,Lee2014,Agterberg2020}. The first example of PDW is the Fulde- Ferrell-Larkin-Ovchinnikov state \cite{FF1964,LO1965} which arises in a conventional $s$-wave superconductor in response to a small degree of spin-polarization so that the Fermi surface is spin-split.
Increasing interest of PDW state has emerged as a mechanism to understand recent discoveries in underdoped cuprate superconductors, where direct observation of PDW has been made experimentally via local Cooper pair tunneling and scanning tunneling microscopy in ${\rm Bi_2Sr_2CaCu_2O_{8+x}}$\cite{Hamidian2016,Ruan2018,Edkins2019} as well as the dynamical inter-layer decoupling observed in $1/8$ hole-doped ${\rm{La}_2BaCuO_4}$ \cite{Berg2007,Agterberg2008}. Although theoretically much is known about the properties of the PDW state, its realization in microscopic models remains still very few.\cite{Berg2010,Fradkin2012,Venderley2019,Xu2019,Han2020,Huang2021} These include the one-dimensional (1D) Kondo-Heisenberg model\cite{Berg2010}, extended two-leg Hubbard-Heisenberg model\cite{Fradkin2012} and an extended Hubbard model with a staggered spin-dependent magnetic flux per plaquette on a three-leg triangular lattice.\cite{Venderley2019}. The signature of the PDW ordering was also observed in $t$-$J$ model with four-spin ring exchange interaction on the four-leg triangular cylinder\cite{Xu2019} and the $t$-$J$-like extension of the Kitaev model on the three-leg honeycomb cylinder.\cite{Cheng2020} However, there is no evidence of the PDW state found in the standard Hubbard model on systems wider than a two-leg ladder.

As a straightforward simplification of quantum chemistry, the single-band Hubbard model has been one of the central paradigms in the field of strongly correlated systems and is widely believed to contain the essential ingredients of high-temperature superconductivity.\cite{Dagotto:1994cz,zhang1988effective,Lee2006,Fradkin2015} Although it has been intensively studied for several decades, new aspects of its rich phase diagram are still regularly unveiled. This is particularly true for the geometrically frustrated triangular lattice where quantum spin liquid (QSL) has been the subject of considerable interest.\cite{Anderson1973,Balents2010}
Encouragingly, a number of experimental evidences suggest that triangular materials $\kappa$-(ET)$_2$Cu$_2$(CN)$_3$\cite{Shimizu2003} and EtMe$_3$Sb[Pd(dmit)$_2$]$_2$\cite{Itou2007,Itou2008,Itou2010,Yamashita2010,Yamashita2011} are promising realization of QSLs.\cite{Senthil2008} Through the substantial theoretical studies of this QSL phase in the context of the Hubbard model and its effective Heisenberg extensions,\cite{Misguich1999,LiMingW2000,Mishmash2013,Kyung2006,ClayRT2008,Morita2002,Koretsune2007,Motrunich2005,LeeSS2005,Yang2010,ShengDN2009,Block2011,Hu2015,QiY2008,Shirakawa2017,Sahebsara2008}, there has been a consensus that the half-filled Hubbard model has a QSL phase at intermediate interaction strength which separates the metallic phase and Mott insulating phase.\cite{Morita2002,Koretsune2007,Motrunich2005,Yang2010,LeeSS2005,Sahebsara2008,Laubach2015,Yoshioka2009,Mizusaki2006,Shirakawa2017,Szasz2020} However, its precise nature remains still under debate: distinct candidate states have been proposed including the QSL with spinon Fermi surfaces\cite{Motrunich2005,Yang2010,LeeSS2005, ShengDN2009,Block2011,Mishmash2015,Shirakawa2017}, $Z_2$ spin liquid\cite{Hu2015,QiY2008} and chiral spin liquid (CSL).\cite{Szasz2020} The debate also exists in the density-matrix renormalization group (DMRG)\cite{White1992} study.\cite{Mishmash2015,Shirakawa2017,Szasz2020,Chen2021} Previous studies suggest a gapless spin liquid in two dimensions (2D)\cite{Shirakawa2017}, however, recent study reports a gapped CSL on both four- and six-leg cylinders.\cite{Szasz2020,Chen2021} To resolve these puzzles, the three-leg triangular cylinder might be an ideal starting point, due to both the essential degrees of freedom accommodating the 2D characteristics and the feasibility of well-controlled DMRG simulations.

Aside from the QSL, a closely related question is the physics of doping it. Intuitively, QSL can be viewed as an insulating phase with preformed electron pairs such that it might produce superconductivity upon light doping.\cite{Anderson1987,Kivelson1987,Rokhsar1988,Laughlin1988,Wen1989,Fradkin2015,Broholm2019} This idea was supported by recent large-scale DMRG studies that nematic $d$-wave,\cite{Jiang2019tqsl} and topological $d\pm id$-wave superconductivity,\cite{Jiang2020tcsl} were observed on the lightly-doped time-reversal symmetric QSL and CSL, respectively. As for the doped Hubbard model on the triangular lattice, a number of SC states are proposed, including the $d$-wave, $d\pm id$-wave, and $p$-wave superconductivity\cite{Raghu2010,ChenKS2013,GuoHM2018,Venderley2019}, however, these were challenged by the recent DMRG study, which reported the absence of superconductivity in doping the Hubbard model on both three- and four-leg cylinders.\cite{Zhu2020}.

\begin{figure}[htbp!]
\centering
    \includegraphics[width=1\linewidth]{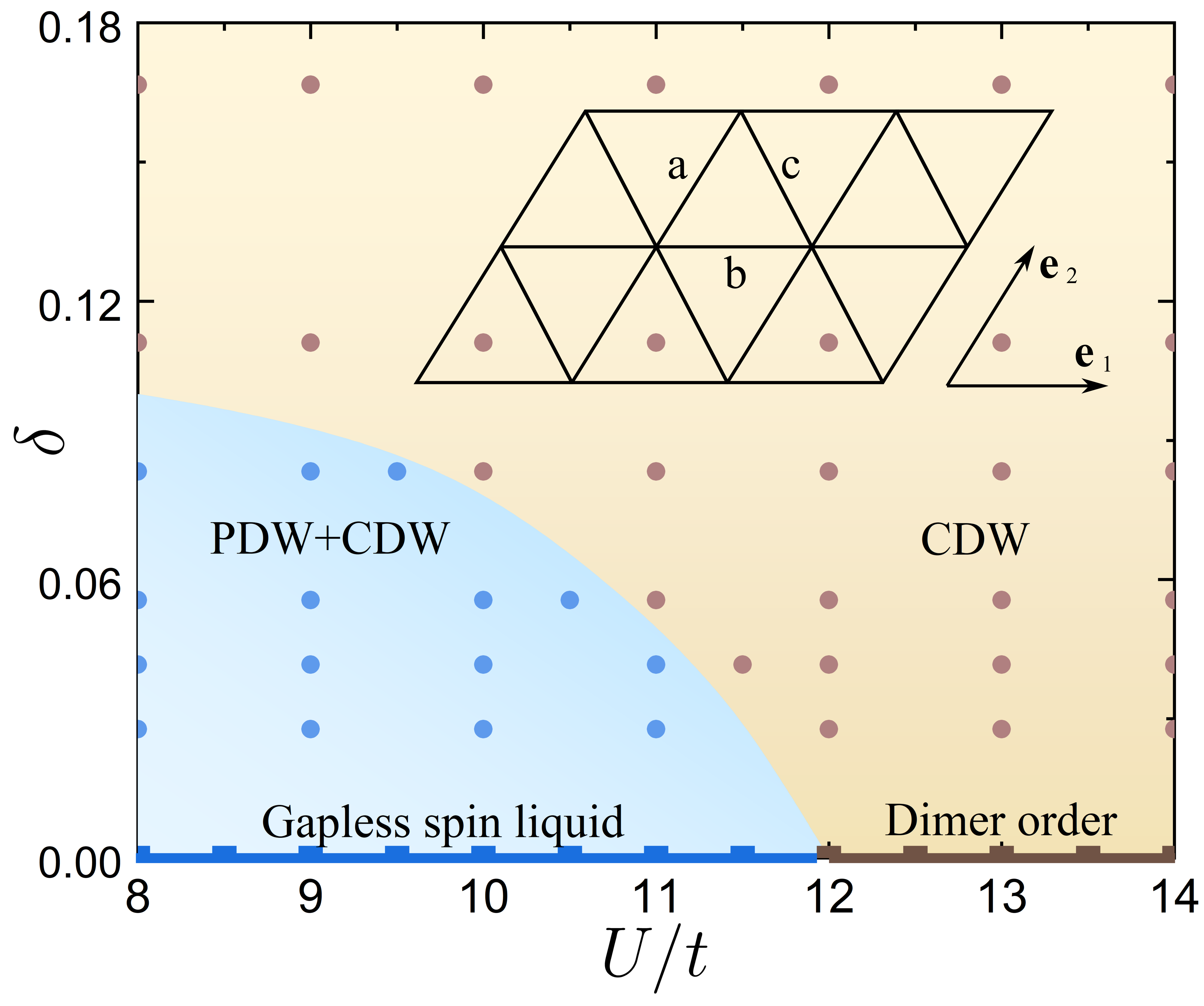}
\caption{(Color online) Ground state phase diagram of the Hubbard model in Eq.(1) on three-leg triangular cylinders as a function of $U/t$ and hole doping concentration $\delta$. The squares are data points for half-filling and the dots are data points at finite doping. Inset: the geometry of the triangular lattice where the two arrows labeled by $\mathbf{e}_1=(1,0)$ and $\mathbf{e}_2=(1/2,\sqrt{3}/2)$ denote the two basis vectors. $a$, $b$ and $c$ label the three different bonds.} \label{Figs:PhaseDiagram}
\end{figure}

\textbf{Principal results: }%
In this paper, we address the above questions by studying the Hubbard model on three-leg triangular cylinders of length up to $L_x=128$ 
using large-scale DMRG simulations. Our main results are summarized in the ground state phase diagram in Fig.\ref{Figs:PhaseDiagram}. At half-filling,  an intermediate time-reversal symmetric gapless 
spin liquid phase separates the metallic phase at weak coupling $U< U_{c1}=7.0\pm 0.5t$ and the Mott insulating dimer phase at strong coupling $U>U_{c2}=12.0\pm 0.5t$. Distinct with the gapped CSL on four- and six-leg cylinders,\cite{Szasz2020,Chen2021} we find that the spin liquid phase on three-leg cylinders is gapless, manifest as gapless spin mode and quasi-long-range spin-spin correlations. With the chiral-chiral correlations decay exponentially at long distances, this phase preserves the time-reversal symmetry. Upon light-doping, this gapless spin liquid evolves into a state consistent with that of the striped PDW\cite{Agterberg2020}. Both the SC correlations and charge-density-wave (CDW) decay as a power-law and oscillate in distance. While other correlations (single-particle, spin-spin, and scalar chiral-chiral) are all short-range, all these correlations are intertwined and mutually commensurate in terms of the wavevector. In contrast, a CDW phase is identified When further doping the gapless spin liquid phase with $\delta\gtrsim 10\%$ or doping the dimer order phase.

\textbf{Model and method: }%
We employ DMRG\cite{White1992} to study the ground-state properties of the Hubbard model on the triangular lattice, whose Hamiltonian is defined as%
\begin{eqnarray}
  H=-t\sum_{\langle ij\rangle \sigma}(\hat{c}^{\dagger}_{i,\sigma}\hat{c}_{j,\sigma}+h.c.) + U\sum_{i}\hat{n}_{i,\uparrow} \hat{n}_{i,\downarrow}.
  \label{Eq:Ham}
\end{eqnarray}
Here, $\hat{c}^{\dagger}_{i\sigma}$ ($\hat{c}_{i\sigma}$) is the electron creation (annihilation) operator with spin-$\sigma$ ($\sigma=\uparrow, \downarrow$) on site $i=(x_i,y_i)$, $\hat{n}_{i,\sigma}=\hat{c}^{\dagger}_{i\sigma}\hat{c}_{i\sigma}$ is the electron number operator. $t$ denotes the electron hopping amplitude between the nearest-neighbor (NN) sites $\langle ij\rangle$, and $U$ is the on-site Coulomb repulsion. The lattice geometry used in our simulations is depicted in the inset of Fig.\ref{Figs:PhaseDiagram}, with open (periodic) boundary condition along the $\mathbf{e}_1$ ($\mathbf{e}_2$) direction. We focus on three-leg triangular cylinders with width $L_y=3$ and length up to $L_x=128$, where $L_y$ and $L_x$ are the number of sites along the $\mathbf{e}_2$ and $\mathbf{e}_1$ directions, respectively. The doped hole concentration is defined as $\delta=N_h/N$, where $N=3L_x$ is the total number of lattice sites and $N_h$ is the number of doped holes. We set $t=1$ as an energy unit and consider $6t\leq U\leq 18t$ in the present study. We perform up to 69 sweeps and keep up to $m=25000$ number of states with a typical truncation error $\epsilon\sim 5\times 10^{-7}$. Further details of the numerical simulation are provided in the Supplemental Material (SM).

\begin{figure}[tb]
\centering
    \includegraphics[width=1\linewidth]{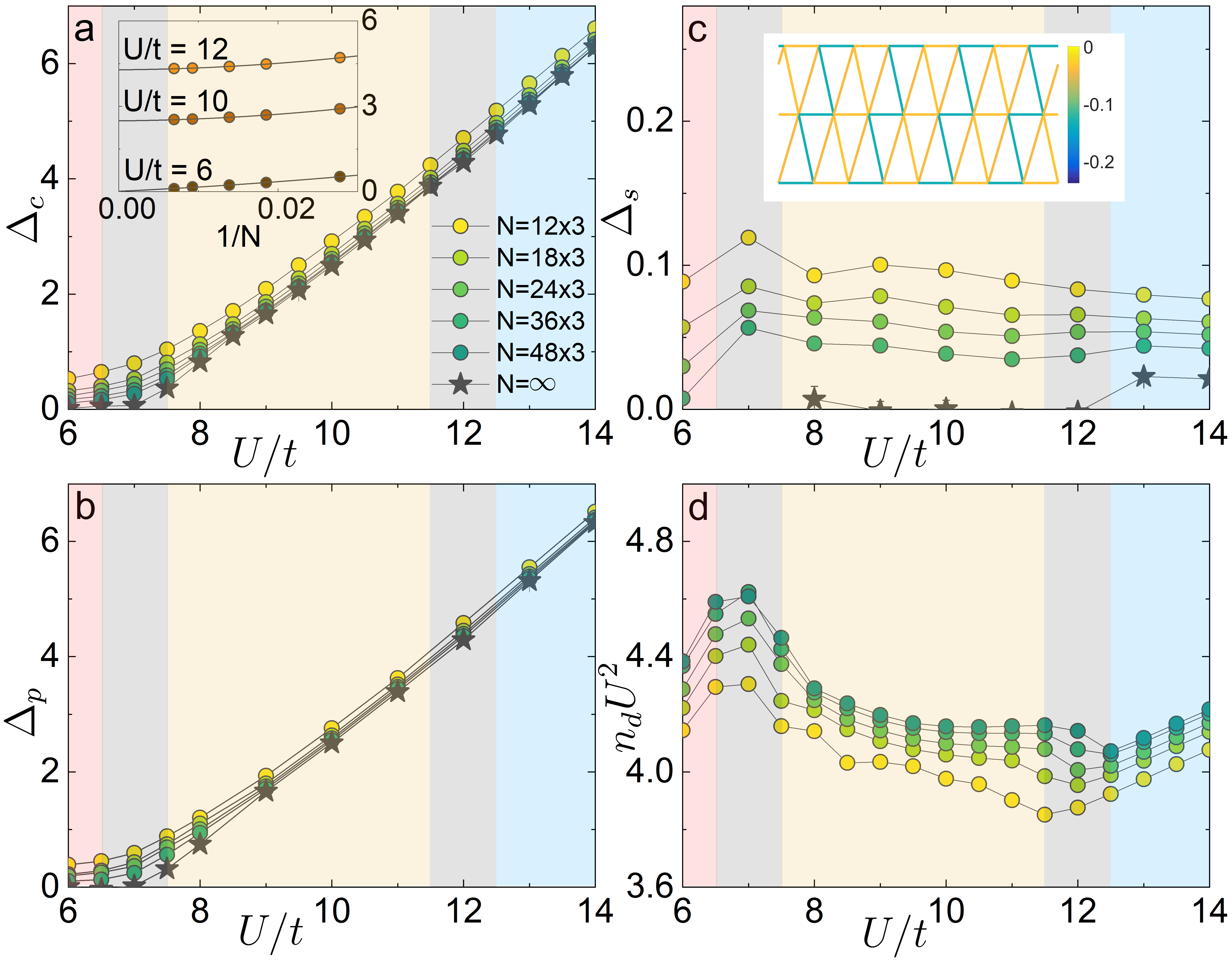}
\caption{(Color online) (a) Charge gap $\Delta_c$, (b) single particle gap $\Delta_p$, (c) spin gap $\Delta_s$, and (d) double occupancy $n_d U^2$ as a function of $U/t$ on three-leg triangular cylinders. The star symbols are extrapolated results in the limit $N\rightarrow \infty$. Insets: (a) Examples of finite-size scaling of $\Delta_c$ at different $U/t$. (c) Dimer pattern, i.e., spin-spin correlation $\langle \mathbf{S}_i\cdot \mathbf{S}_j\rangle$ between NN sites $\langle ij\rangle$ at $U/t= 15$. The grey shaded regions denote the phase boundaries at $U_{c1}$ and $U_{c2}$.} \label{Figs:PBoundary}
\end{figure}

\begin{table*}[tb]
\centering 
\begin{tabular}{ c   C{0.12\linewidth}   C{0.06\linewidth}   C{0.06\linewidth}   C{0.06\linewidth}   C{0.06\linewidth}   C{0.06\linewidth}   C{0.07\linewidth}  C{0.07\linewidth} C{0.06\linewidth} C{0.06\linewidth}} 
\hline\hline 
Parameters & Phase & $K_s$ & $\xi_{s}$ & $\xi_{G}$ &  $\xi_{\chi}$ & $K_c$ & $K_{sc}$($aa$) & $K_{sc}$($cc$) & $\xi_{sc}$ & $c$ \\ [.5ex] 
\hline 
 $U=10t, \delta=0$, $L_x\leq 72$ & Gapless QSL & $1.1(1)$ & -- & $1.1(2)$ & $2.2(1)$ & -- & -- & -- & -- & $\sim 1$   \\
 $U=9t, \delta=1/18$, $L_x\leq 108$ & PDW+CDW & -- & $15(1)$ & $4.5(1)$ & $6.2(1)$ & $1.6(1)$ & $3.6(2)$ & $3.9(3)$ & -- & --  \\
 $U=10t, \delta=1/24$, $L_x\leq 128$ & PDW+CDW & -- & $22(1)$ & $4.1(1)$ & $5.5(1)$ & $1.6(1)$ & $3.6(2)$ & $3.9(3)$ & -- & --  \\
 $U=18t, \delta=1/18$, $L_x\leq 72$ & CDW & -- & $5.9(1)$ & $10.8(5)$ & $5.4(2)$ & $1.6(1)$ & -- & -- & $8.3(1)$ & $\sim 1$  \\
\hline\hline 
\end{tabular}
\caption{Summary of the phases. Parameters, corresponding phases, exponents ($K_s$, $K_c$, $K_{sc}$), correlation lengths ($\xi_s$, $\xi_G$, $\xi_\chi$) and central charge $c$. Note that $K_c$ shown in the table is determined from the Friedel oscillation, and $K_{sc}$ is extracted from SC correlation $\Phi_{aa(cc)}(r)$. The cylinder lengths and correlation lengths are in the unit of lattice spacing.}\label{Table:IntMidQSExponent}
\end{table*}

\textbf{Gapless spin liquid: }%
At half-filling, we identify three distinct phases (see Fig.\ref{Figs:PhaseDiagram} and Fig.\ref{Figs:PBoundary}) separated by two phase transitions at $U_{c1}=7.0\pm 0.5t$ and $U_{c2}=12.0\pm 0.5t$. These phases are determined by various energy gaps including the single-particle gap $\Delta_p$, charge gap $\Delta_c$ and spin-triplet gap $\Delta_s$ defined as%
\begin{eqnarray}
\Delta_{p} &=& E_{\frac{N}{2}+1,\frac{N}{2}}+E_{\frac{N}{2}-1,\frac{N}{2}}-2E_{\frac{N}{2},\frac{N}{2}}, \nonumber \\%
\Delta_{c} &=& [E_{\frac{N}{2}+1,\frac{N}{2}+1}+E_{\frac{N}{2}-1,\frac{N}{2}-1}-2E_{\frac{N}{2},\frac{N}{2}}]/2, \\%
\Delta_s &=& E_{\frac{N}{2}+1,\frac{N}{2}-1}-E_{\frac{N}{2},\frac{N}{2}}. \nonumber
\label{Eq:Gap}
\end{eqnarray}
Here $E_{N_{\uparrow},N_{\downarrow}}$ is the ground state energy of the system with $N_{\uparrow}$ spin-up and $N_{\downarrow}$ spin-down electrons. Our calculations identify a metallic phase at $U<U_{c1}$ where all three gaps vanish in the thermodynamic limit, consistent with previous studies\,\cite{Yang2010,Mizusaki2006,Shirakawa2017,Szasz2020}. At large $U>U_{c2}$, the ground state of the system can be mapped onto the spin-$1/2$ antiferromagnetic Heisenberg model. It has a dimerized ground state on three-leg cylinders\cite{Ru2013} where all three gaps are expected to be finite in the thermodynamic limit. This is indeed consistent with our results as shown in Fig.\ref{Figs:PBoundary}a-c including the dimer pattern in the inset of Fig.\ref{Figs:PBoundary}c. Independently, the phase boundaries can also be determined by $n_d U^2$, with the double occupancy $n_d=\frac{1}{N}\sum_i\langle \hat{n}_{i,\uparrow} \hat{n}_{i,\downarrow}\rangle$\cite{Yang2010}, which exhibits peak and kink at the two phase boundaries (see Fig.\ref{Figs:PBoundary}d).

We focus on the intermediate phase among these three phases.
Distinct with four- and six-leg cylinders, this intermediate phase on three-leg cylinders is consistent with a gapless spin liquid, where both $\Delta_p$ and $\Delta_c$ remain finite but $\Delta_s$ vanishes in the thermodynamic limit as shown in Fig.\ref{Figs:PBoundary}a-c. To further support this, we consider $U=10t$ as an example (deeply in the intermediate phase) and investigate the scaling behavior. 

\begin{figure}[tb]
\centering
    \includegraphics[width=1\linewidth]{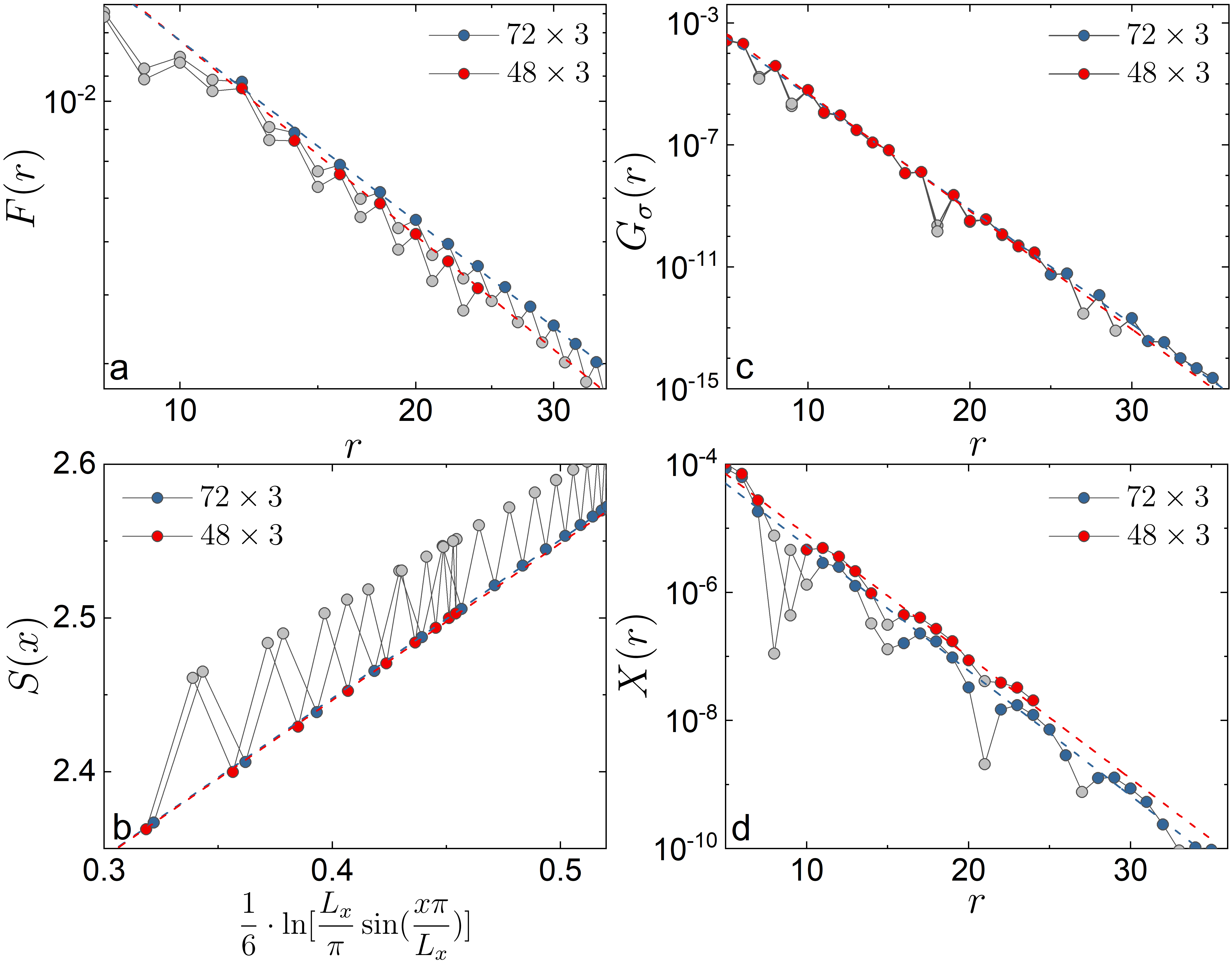}
\caption{(Color online) Correlation functions of the Hubbard model at half-filling with $U=10t$. Data points in gray color are discarded in fittings. (a) Spin-spin correlation $F(r)$ and its power-law fit $f(r)\sim r^{-K_s}$ labelled by the dashed lines. (b) Von Neumann entanglement entropy $S(x)$ where the slope of the dashed lines gives the central charge $c$. (c) Single-particle correlation $G_{\sigma}(r)$ and its exponential fit $f(r)\sim e^{-r/\xi_{G}}$ labelled by the dashed lines. (d) Scalar chiral-chiral correlation $X(r)$ and its exponential fit $f(r)\sim e^{-r/\xi_{\chi}}$ labelled by the dashed lines.}\label{Figs:HFU10CS}
\end{figure}

We first calculate the spin-spin correlation%
\begin{eqnarray}\label{Eq:SpinCor}
F(r) = \frac{1}{L_y}\sum_{y=1}^{L_y}|\langle \mathbf{S}_{(x_0,y)}\cdot \mathbf{S}_{(x_0+r,y)}\rangle|,
\end{eqnarray}
where $\mathbf{S}_{i}$ is the $S=1/2$ spin operator on site $i$ and ($x_0,y$) is the reference site with $x_0\sim L_x/4$ and $r$ is the distance between two sites in the $\mathbf{e}_1$ direction. As shown in Fig.\ref{Figs:HFU10CS}a, it is clear that $F(r)$ decays with a power-law at long distances which can be well fitted by $F(r)\sim r^{-K_s}$ with corresponding Luttinger exponent $K_s=1.1(1)$. 
As a further test, a key feature of the gapless spin liquid is its finite gapless spin mode characterized by the central charge $c$. It can be obtained from fitting the von Neumann entanglement entropy, $S(x)=-{\rm Tr}[\rho_x \ln \rho_x]$, through
$S(x) = \frac{c}{6}\ln[\frac{L_x}{\pi}\sin(\frac{\pi x}{L_x})] + {\rm const}$, where $\rho_x$ is the reduced density matrix of a (quasi-) 1D subsystem with length $x$\,\cite{Calabrese2004,Fagotti2011}. For critical (quasi-) 1D systems, it has been established\cite{Calabrese2004,Fagotti2011} that $S(x) = \frac{c}{6}\ln[\frac{L_x}{\pi}\sin(\frac{\pi x}{L_x})] + {\rm const}$. Examples are shown in Fig.\ref{Figs:HFU10CS}b for cylinders of length $L_x=48$ and $L_x=72$, the extracted central charge is $c=1.0(1)$ suggesting that the intermediate phase has one gapless mode.

In contrast to the spin channel, a finite single-particle gap in the intermediate phase suggests that the single-particle correlation%
\begin{eqnarray}\label{Eq:GreenFunc}
  G_\sigma(r)=\frac{1}{L_y}\sum_{y=1}^{L_y}\langle c^{\dagger}_{(x_0,y),\sigma} c_{(x_0+r,y),\sigma}\rangle,
\end{eqnarray}
should decay exponentially as $G_\sigma(r)\sim e^{-r/\xi_G}$ with a correlation length $\xi_G$. This is indeed the case as shown in Fig.\ref{Figs:HFU10CS}c, where $G_\sigma(r)$ decays exponentially and the extracted correlation length is $\xi_G=1.1(2)$.

To test the possibility of time-reversal symmetry breaking, we have also calculated the scalar chiral-chiral correlation $X(r)$, which is defined as%
\begin{equation}\label{Eq:ChiralCor}
     X(r)=\frac{1}{L_y}\sum_{y=1}^{L_y}|\langle \chi_{(x_0,y)} \chi_{(x_0+r,y)}\rangle|.
\end{equation}
Here $\chi_{i}=\mathbf{S}_i \cdot (\mathbf{S}_j \times \mathbf{S}_k)$ is the scalar chiral operator, where $i$, $j$ and $k$ label clockwise vertices of a triangle. On three-leg cylinders, we find that $X(r)$ decays exponentially as $X(r)\sim e^{-r/\xi_\chi}$ at long distances with the correlation length $\xi_\chi=2.2(1)$. Therefore, we conclude that the intermediate gapless spin liquid phase on three-leg cylinders preserves time-reversal symmetry, in stark contrast to the gapped CSL on four- and six-leg cylinders.\cite{Szasz2020}

\begin{figure}[!th]
\centering
    \includegraphics[width=1\linewidth]{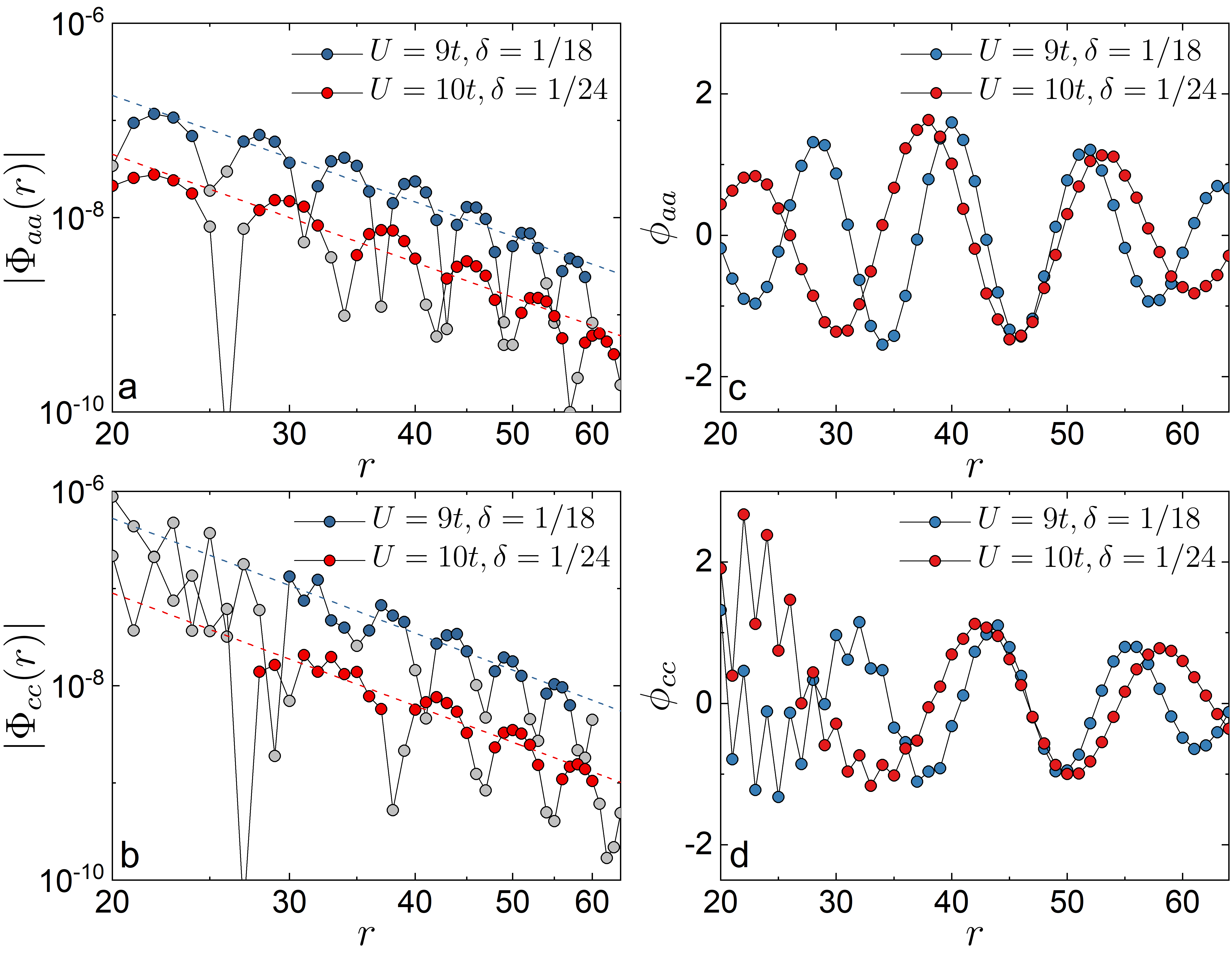}
\caption{(Color online) Superconducting correlations (a) $|\Phi_{aa}(r)|$ and (b) $|\Phi_{cc}(r)|$ where dashed lines denote fittings to a power-law function $f(r)\sim r^{-K_{sc}}$. Data points in gray color are discarded in the fitting. The normalized functions (c) $\phi_{aa}(r)=(-1)^r \Phi_{aa}(r)/f(r)$ and (d) $\phi_{cc}(r)=(-1)^r \Phi_{cc}(r)/f(r)$ reflect the spatial oscillation of $\Phi_{aa}(r)$ and $\Phi_{cc}(r)$, respectively.}\label{Figs:PDW_SC}
\end{figure}

\textbf{Lightly doped gapless spin liquid: }%
Upon light doping the gapless spin liquid, a state which is consistent with that of the striped PDW emerges where the CDW and SC pair-field correlations decay spatially in a power-law at long distances. We provide two detailed examples ($U=9t$, $\delta=1/18$ and $U=10t$, $\delta=1/24$) in Fig.\ref{Figs:PDW_SC}, while the conclusion holds for all parameter in the PDW+CDW phase of Fig.~\ref{Figs:PhaseDiagram}. In this paper, we have studied a sizeable system with length up to $L_x=128$ to suppress the finite-size effect. As shown below, the oscillation period of SC correlations is rather large, which results in the absence of the PDW correlation in previous study.\cite{Zhu2020}

\textit{Pair density wave. }%
To test the possibility of superconductivity, we calculate the equal-time SC pair-field correlations. As the ground state with an even number of electrons always have total spin 0, we focus on spin-singlet SC correlation, which is defined as%
\begin{eqnarray}
\Phi_{\alpha\beta}(r)=\frac{1}{L_y}\sum_{y=1}^{L_y}|\langle\Delta^{\dagger}_{\alpha}(x_0,y)\Delta_{\beta}(x_0+r,y)\rangle|. \label{Eq:def_SC_cor}
\end{eqnarray}
Here, $\Delta^{\dagger}_{\alpha}(x,y)=\frac{1}{\sqrt{2}}[\hat{c}^{\dagger}_{(x,y),\uparrow}\hat{c}^{\dagger}_{(x,y)+\alpha,\downarrow}-\hat{c}^{\dagger}_{(x,y),\downarrow}\hat{c}^{\dagger}_{(x,y)+\alpha,\uparrow}]$ is spin-singlet pair creation operator living on bond $\alpha=$a, b and c (see Fig.\ref{Figs:PhaseDiagram} inset). ($x_0,y$) is the reference site with $x_0\sim L_x/4$ and $r$ is the distance between two bonds in the $\mathbf{e}_1$ direction. The spatial distribution of SC correlations $\Phi_{aa}(r)$ and $\Phi_{cc}(r)$ for the two examples are shown in  Fig.\ref{Figs:PDW_SC}: $\Phi(r)$ exhibits clear spatial oscillation which can be well fitted by $\Phi(r)\sim f(r)\phi(r)$ for a large region of $r$, where $f(r)$ sets envelope and $\phi(r)$ determines the oscillation, as discussed below. At long distances, the envelope function $f(r)$ is consistent with a power-law decay $f(r)\sim r^{-K_{sc}}$. The extracted exponent is $K_{sc}=3.6(2)$ for $\Phi_{aa}(r)$ and $K_{sc}=3.9(3)$ for $\Phi_{cc}(r)$, respectively. We have also calculated the spin-triplet SC correlations, which however are much weaker than the spin-singlet SC correlations.

The spatial oscillation of the SC correlations $\Phi(r)$ is characterized by the normalized function $\phi(r)=(-1)^r\Phi(r)/f(r)$ as mentioned above. Examples of $\phi_{aa}(r)$ and $\phi_{cc}(r)$ are shown in Fig.\ref{Figs:PDW_SC}c-d, both of which oscillate periodically in real space and can be well fitted by $\phi(r)\sim \sin (Qr +\theta)$ for $\phi_{aa}(r)$ when $r\gtrsim 8$ and $\phi_{cc}(r)$ when $r\gtrsim 24$. This is consistent with the striped PDW state with vanishing spatial average of $\phi(r)$. $Q=3\pi\delta$ is the corresponding PDW ordering wavevector which corresponds to the wavelength $\lambda_{sc}=2/3\delta$, i.e., $\lambda_{sc}=12$ for $\delta=1/18$ and $\lambda_{sc}=16$ for $\delta=1/24$. As we will see below, our results clearly show the relationship $\lambda_{sc}=\lambda_s=2\lambda_c=\lambda_{\chi}$, which is expected for the striped PDW state. Here $\lambda_s$, $\lambda_c$ and $\lambda_{\chi}$ are wavelengths of the spin-spin, CDW and scalar chiral-chiral correlations.

\begin{figure}[htb]
\centering
    \includegraphics[width=1\linewidth]{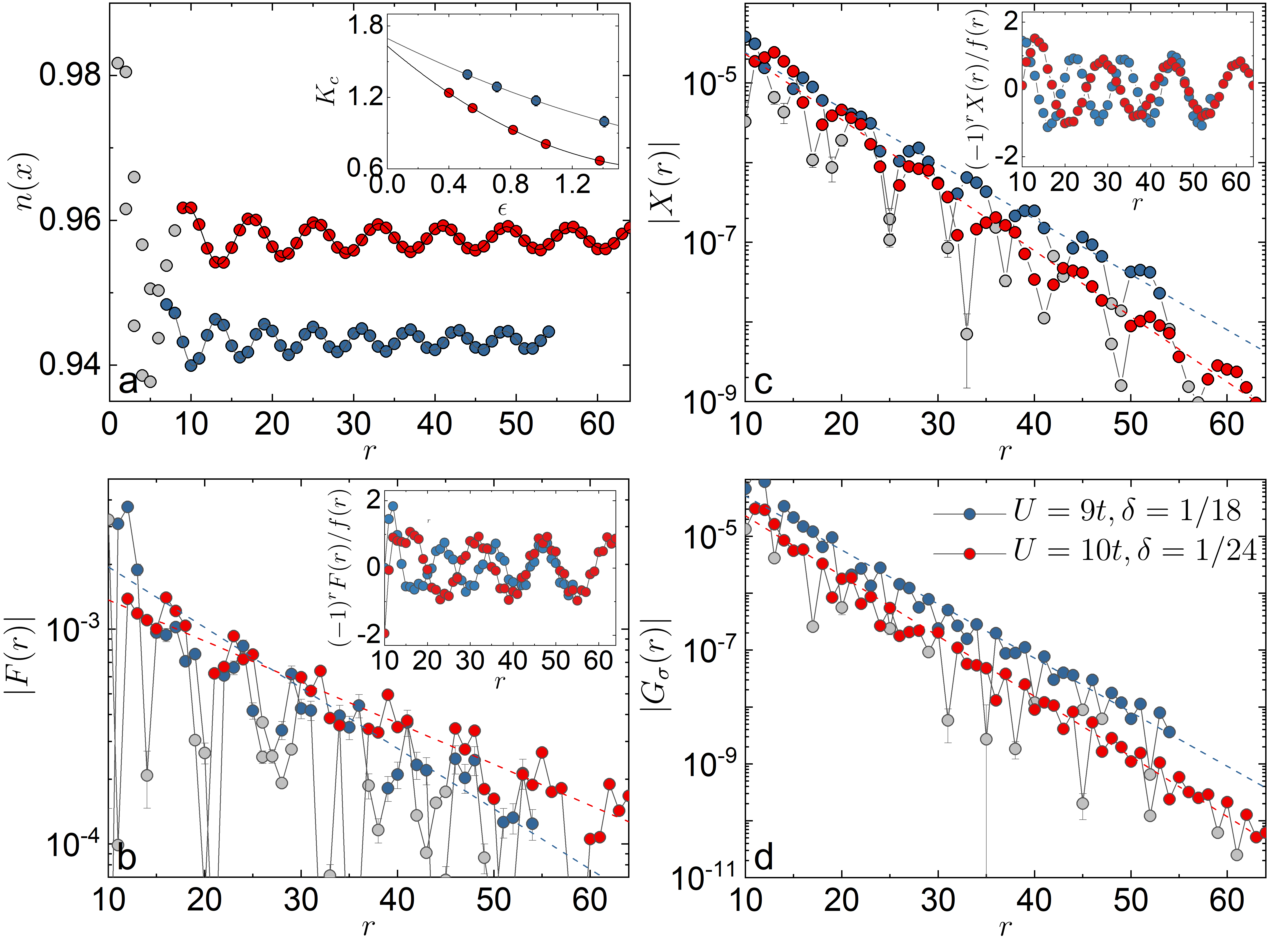}
\caption{(Color online) (a) Charge density profile $n(x)$ where the solid lines denote the fitting using Eq.(\ref{Eq:Friedel}). Data points in gray are discarded to minimize the boundary effect. Inset is the extracted exponent $K_c$ as a function of truncation error $\epsilon$. (b) Spin-spin correlation $F(r)$ and the exponential fitting $f(r)\sim e^{-r/\xi_{s}}$ (dashed lines). Inset: Normalized correlation $(-1)^r F(r)/f(r)$. (c) Chiral-chiral correlation $|X(r)|$ and the exponential fitting $f(r)\sim e^{-r/\xi_\chi}$ (dashed lines). Inset: Normalized correlation $(-1)^r X(r)/f(r)$. (d) Single-particle correlation $G_{\sigma}(r)$ and the exponential fitting $f(r)\sim e^{-r/\xi_G}$ labelled by the dashed lines.}\label{Figs:PDW_CorlFunc}
\end{figure}

\textit{Charge density wave. }%
To measure the charge order, we define the local rung density operator as $\hat{n}(x)=\sum_{y=1}^{L_y}\hat{n}(x,y)$ and its expectation value as $n(x)=\langle \hat{n}(x)\rangle$. Fig.\ref{Figs:PDW_CorlFunc}a shows the charge density profile $n(x)$ on cylinders of length $L_x=108$ at $\delta=1/18$ and $L_x=128$ at $\delta=1/24$. The system forms $1/3$-filled charge stripes of wavelength $\lambda_c=1/3\delta$, which is the spacing between two adjacent charge stripe along the cylinder. This corresponds to an ordering wavevector $K=6\pi\delta=2Q$ with 1/3 doped hole per CDW unit cell.

At long distances, the spatial decay of CDW correlations is dominated by a power-law with the Luttinger exponent $K_c$, which can be obtained by fitting the charge density oscillations (Friedel oscillations) induced by the boundaries of the cylinder\cite{White2002,Moreno2011}%
\begin{eqnarray}
n(x)=n_0+\delta n \cos(K*x+\theta)x^{-K_c/2}. \label{Eq:Friedel}
\end{eqnarray}
Here $n_0$ denotes the background electron density, $\delta n$ and $\theta$ are model-dependent constants. Note that a few data points (Fig.\ref{Figs:PDW_CorlFunc}a, in gray) are excluded to minimize the boundary effect for a more reliable fit. The extracted exponent $K_c=1.6(1)$ is shown in the inset of Fig.\ref{Figs:PDW_CorlFunc}a. Alternatively, $K_c$ can also be obtained from the charge density-density correlation, which gives consistent results (see SM for details).

\textit{Other correlations. }%
To further characterize the PDW phase, we have also calculated other correlations including $F(r)$, $X(r)$ and $G_\sigma(r)$ as shown in Fig.\ref{Figs:PDW_CorlFunc}. Contrary to CDW and SC correlations, we find that they decay exponentially at long distances as $F(r)\sim e^{-r/\xi_s}$, $X(r)\sim e^{-r/\xi_\chi}$ and $G_\sigma(r)\sim e^{-r/\xi_G}$, where the corresponding correlation lengths $\xi_s$, $\xi_\chi$ and $\xi_G$ are given in Table \ref{Table:IntMidQSExponent}. It may be worth mentioning that while $F(r)$ decays exponentially at long distances, its correlation length is fairly long $\xi_s\sim 22(2)$. This can be attributed to the fact that the lightly doped case is very close to the gapless spin liquid at half-filling, which has divergent correlation length. Interestingly, we find that both $F(r)$ and $X(r)$ exhibit clear spatial oscillation as shown in the insets of Fig.\ref{Figs:PDW_CorlFunc}b-c with wavelengths $\lambda_s$ and $\lambda_\chi$ that are the same as that of the SC correlation, i.e., $\lambda_s=\lambda_\chi=\lambda_{sc}$. This gives the same ordering wavevector $Q$ as the SC correlation. These features further support the striped PDW state in the lightly doped system.

\textbf{Conclusion: }%
In summary, we have studied the ground state properties of the Hubbard model on sizeable three-leg triangular cylinders. Based on our results, we conclude that the exact ground state of the system has the following properties: (1) At half-filling, there is an intermediate gapless spin liquid phase which is characterized by one gapless spin mode and power-law spin-spin correlation but a gap to all charge excitations. (2) Light doping ($\delta\lesssim 10\%$) the gapless spin liquid phase can give rise to a striped PDW state with power-law SC correlations with moderate exponent $K_{sc}\sim 4$ and an ordering wavevector $Q$. (3) There are power-law CDW correlations with an ordering wavevector $K=2Q$. (4) While both spin-spin and scalar chiral-chiral correlations are short-ranged, they are mutually commensurate with both CDW and SC correlations with an ordering wavevector $Q$. To the best of our knowledge, this is the first numerical observation of power-law PDW correlation in the standard Hubbard model on a system wider than the 2-leg ladder. 

{\it Acknowledgments:} We would like to thank Thomas Devereaux and especially Steve Kivelson for insightful discussion and invaluable suggestions. This work was supported by the Department of Energy, Office of Science, Basic Energy Sciences, Materials Sciences and Engineering Division, under Contract DE-AC02-76SF00515. Y.W. acknowledges support from National Science Foundation (NSF) award DMR-2038011. Parts of the computing for this project was performed on the Sherlock cluster. Parts of the calculations in Fig.\ref{Figs:PhaseDiagram} and Fig.\ref{Figs:PBoundary} are performed using the high-performance matrix product state algorithm library GraceQ/MPS2\cite{GraceQ}

\bibliography{references}

\begin{thebibliography}{70}%
\makeatletter
\providecommand \@ifxundefined [1]{%
 \@ifx{#1\undefined}
}%
\providecommand \@ifnum [1]{%
 \ifnum #1\expandafter \@firstoftwo
 \else \expandafter \@secondoftwo
 \fi
}%
\providecommand \@ifx [1]{%
 \ifx #1\expandafter \@firstoftwo
 \else \expandafter \@secondoftwo
 \fi
}%
\providecommand \natexlab [1]{#1}%
\providecommand \enquote  [1]{``#1''}%
\providecommand \bibnamefont  [1]{#1}%
\providecommand \bibfnamefont [1]{#1}%
\providecommand \citenamefont [1]{#1}%
\providecommand \href@noop [0]{\@secondoftwo}%
\providecommand \href [0]{\begingroup \@sanitize@url \@href}%
\providecommand \@href[1]{\@@startlink{#1}\@@href}%
\providecommand \@@href[1]{\endgroup#1\@@endlink}%
\providecommand \@sanitize@url [0]{\catcode `\\12\catcode `\$12\catcode
  `\&12\catcode `\#12\catcode `\^12\catcode `\_12\catcode `\%12\relax}%
\providecommand \@@startlink[1]{}%
\providecommand \@@endlink[0]{}%
\providecommand \url  [0]{\begingroup\@sanitize@url \@url }%
\providecommand \@url [1]{\endgroup\@href {#1}{\urlprefix }}%
\providecommand \urlprefix  [0]{URL }%
\providecommand \Eprint [0]{\href }%
\providecommand \doibase [0]{http://dx.doi.org/}%
\providecommand \selectlanguage [0]{\@gobble}%
\providecommand \bibinfo  [0]{\@secondoftwo}%
\providecommand \bibfield  [0]{\@secondoftwo}%
\providecommand \translation [1]{[#1]}%
\providecommand \BibitemOpen [0]{}%
\providecommand \bibitemStop [0]{}%
\providecommand \bibitemNoStop [0]{.\EOS\space}%
\providecommand \EOS [0]{\spacefactor3000\relax}%
\providecommand \BibitemShut  [1]{\csname bibitem#1\endcsname}%
\let\auto@bib@innerbib\@empty
\bibitem [{\citenamefont {Lee}\ and\ \citenamefont {Lee}(2005)}]{LeeSS2005}%
  \BibitemOpen
  \bibfield  {author} {\bibinfo {author} {\bibfnamefont {S.-S.}\ \bibnamefont
  {Lee}}\ and\ \bibinfo {author} {\bibfnamefont {P.~A.}\ \bibnamefont {Lee}},\
  }\href {\doibase 10.1103/PhysRevLett.95.036403} {\bibfield  {journal}
  {\bibinfo  {journal} {Phys. Rev. Lett.}\ }\textbf {\bibinfo {volume} {95}},\
  \bibinfo {pages} {036403} (\bibinfo {year} {2005})}\BibitemShut {NoStop}%
\bibitem [{\citenamefont {Lee}(2014)}]{Lee2014}%
  \BibitemOpen
  \bibfield  {author} {\bibinfo {author} {\bibfnamefont {P.~A.}\ \bibnamefont
  {Lee}},\ }\href {\doibase 10.1103/PhysRevX.4.031017} {\bibfield  {journal}
  {\bibinfo  {journal} {Phys. Rev. X}\ }\textbf {\bibinfo {volume} {4}},\
  \bibinfo {pages} {031017} (\bibinfo {year} {2014})}\BibitemShut {NoStop}%
\bibitem [{\citenamefont {Agterberg}\ \emph {et~al.}(2020)\citenamefont
  {Agterberg}, \citenamefont {Davis}, \citenamefont {Edkins}, \citenamefont
  {Fradkin}, \citenamefont {Van~Harlingen}, \citenamefont {Kivelson},
  \citenamefont {Lee}, \citenamefont {Radzihovsky}, \citenamefont {Tranquada},\
  and\ \citenamefont {Wang}}]{Agterberg2020}%
  \BibitemOpen
  \bibfield  {author} {\bibinfo {author} {\bibfnamefont {D.~F.}\ \bibnamefont
  {Agterberg}}, \bibinfo {author} {\bibfnamefont {J.~S.}\ \bibnamefont
  {Davis}}, \bibinfo {author} {\bibfnamefont {S.~D.}\ \bibnamefont {Edkins}},
  \bibinfo {author} {\bibfnamefont {E.}~\bibnamefont {Fradkin}}, \bibinfo
  {author} {\bibfnamefont {D.~J.}\ \bibnamefont {Van~Harlingen}}, \bibinfo
  {author} {\bibfnamefont {S.~A.}\ \bibnamefont {Kivelson}}, \bibinfo {author}
  {\bibfnamefont {P.~A.}\ \bibnamefont {Lee}}, \bibinfo {author} {\bibfnamefont
  {L.}~\bibnamefont {Radzihovsky}}, \bibinfo {author} {\bibfnamefont {J.~M.}\
  \bibnamefont {Tranquada}}, \ and\ \bibinfo {author} {\bibfnamefont
  {Y.}~\bibnamefont {Wang}},\ }\href {\doibase
  10.1146/annurev-conmatphys-031119-050711} {\bibfield  {journal} {\bibinfo
  {journal} {Annual Review of Condensed Matter Physics}\ }\textbf {\bibinfo
  {volume} {11}},\ \bibinfo {pages} {231} (\bibinfo {year} {2020})}\BibitemShut
  {NoStop}%
\bibitem [{\citenamefont {Fulde}\ and\ \citenamefont {Ferrell}(1964)}]{FF1964}%
  \BibitemOpen
  \bibfield  {author} {\bibinfo {author} {\bibfnamefont {P.}~\bibnamefont
  {Fulde}}\ and\ \bibinfo {author} {\bibfnamefont {R.~A.}\ \bibnamefont
  {Ferrell}},\ }\href {\doibase 10.1103/PhysRev.135.A550} {\bibfield  {journal}
  {\bibinfo  {journal} {Phys. Rev.}\ }\textbf {\bibinfo {volume} {135}},\
  \bibinfo {pages} {A550} (\bibinfo {year} {1964})}\BibitemShut {NoStop}%
\bibitem [{\citenamefont {Larkin}\ and\ \citenamefont
  {Ovchinnikov}(1965)}]{LO1965}%
  \BibitemOpen
  \bibfield  {author} {\bibinfo {author} {\bibfnamefont {A.~I.}\ \bibnamefont
  {Larkin}}\ and\ \bibinfo {author} {\bibfnamefont {Y.~N.}\ \bibnamefont
  {Ovchinnikov}},\ }\href@noop {} {\bibfield  {journal} {\bibinfo  {journal}
  {Sov. Phys. JETP}\ }\textbf {\bibinfo {volume} {20}},\ \bibinfo {pages} {762}
  (\bibinfo {year} {1965})}\BibitemShut {NoStop}%
\bibitem [{\citenamefont {Hamidian}\ \emph {et~al.}(2016)\citenamefont
  {Hamidian}, \citenamefont {Edkins}, \citenamefont {Joo}, \citenamefont
  {Kostin}, \citenamefont {Eisaki}, \citenamefont {Uchida}, \citenamefont
  {Lawler}, \citenamefont {Kim}, \citenamefont {Mackenzie}, \citenamefont
  {Fujita}, \citenamefont {Lee},\ and\ \citenamefont {Davis}}]{Hamidian2016}%
  \BibitemOpen
  \bibfield  {author} {\bibinfo {author} {\bibfnamefont {M.~H.}\ \bibnamefont
  {Hamidian}}, \bibinfo {author} {\bibfnamefont {S.~D.}\ \bibnamefont
  {Edkins}}, \bibinfo {author} {\bibfnamefont {S.~H.}\ \bibnamefont {Joo}},
  \bibinfo {author} {\bibfnamefont {A.}~\bibnamefont {Kostin}}, \bibinfo
  {author} {\bibfnamefont {H.}~\bibnamefont {Eisaki}}, \bibinfo {author}
  {\bibfnamefont {S.}~\bibnamefont {Uchida}}, \bibinfo {author} {\bibfnamefont
  {M.~J.}\ \bibnamefont {Lawler}}, \bibinfo {author} {\bibfnamefont {E.-A.}\
  \bibnamefont {Kim}}, \bibinfo {author} {\bibfnamefont {A.~P.}\ \bibnamefont
  {Mackenzie}}, \bibinfo {author} {\bibfnamefont {K.}~\bibnamefont {Fujita}},
  \bibinfo {author} {\bibfnamefont {J.}~\bibnamefont {Lee}}, \ and\ \bibinfo
  {author} {\bibfnamefont {J.~C.~S.}\ \bibnamefont {Davis}},\ }\href@noop {}
  {\bibfield  {journal} {\bibinfo  {journal} {Nature}\ }\textbf {\bibinfo
  {volume} {532}},\ \bibinfo {pages} {343} (\bibinfo {year}
  {2016})}\BibitemShut {NoStop}%
\bibitem [{\citenamefont {Ruan}\ \emph {et~al.}(2018)\citenamefont {Ruan},
  \citenamefont {Li}, \citenamefont {Hu}, \citenamefont {Hao}, \citenamefont
  {Li}, \citenamefont {Cai}, \citenamefont {Zhou}, \citenamefont {Lee},\ and\
  \citenamefont {Wang}}]{Ruan2018}%
  \BibitemOpen
  \bibfield  {author} {\bibinfo {author} {\bibfnamefont {W.}~\bibnamefont
  {Ruan}}, \bibinfo {author} {\bibfnamefont {X.}~\bibnamefont {Li}}, \bibinfo
  {author} {\bibfnamefont {C.}~\bibnamefont {Hu}}, \bibinfo {author}
  {\bibfnamefont {Z.}~\bibnamefont {Hao}}, \bibinfo {author} {\bibfnamefont
  {H.}~\bibnamefont {Li}}, \bibinfo {author} {\bibfnamefont {P.}~\bibnamefont
  {Cai}}, \bibinfo {author} {\bibfnamefont {X.}~\bibnamefont {Zhou}}, \bibinfo
  {author} {\bibfnamefont {D.-H.}\ \bibnamefont {Lee}}, \ and\ \bibinfo
  {author} {\bibfnamefont {Y.}~\bibnamefont {Wang}},\ }\href {\doibase
  10.1038/s41567-018-0276-8} {\bibfield  {journal} {\bibinfo  {journal} {Nature
  Physics}\ ,\ \bibinfo {pages} {1178}} (\bibinfo {year} {2018})}\BibitemShut
  {NoStop}%
\bibitem [{\citenamefont {Edkins}\ \emph {et~al.}(2019)\citenamefont {Edkins},
  \citenamefont {Kostin}, \citenamefont {Fujita}, \citenamefont {Mackenzie},
  \citenamefont {Eisaki}, \citenamefont {Uchida}, \citenamefont {Sachdev},
  \citenamefont {Lawler}, \citenamefont {Kim}, \citenamefont
  {S{\'e}amus~Davis},\ and\ \citenamefont {Hamidian}}]{Edkins2019}%
  \BibitemOpen
  \bibfield  {author} {\bibinfo {author} {\bibfnamefont {S.~D.}\ \bibnamefont
  {Edkins}}, \bibinfo {author} {\bibfnamefont {A.}~\bibnamefont {Kostin}},
  \bibinfo {author} {\bibfnamefont {K.}~\bibnamefont {Fujita}}, \bibinfo
  {author} {\bibfnamefont {A.~P.}\ \bibnamefont {Mackenzie}}, \bibinfo {author}
  {\bibfnamefont {H.}~\bibnamefont {Eisaki}}, \bibinfo {author} {\bibfnamefont
  {S.}~\bibnamefont {Uchida}}, \bibinfo {author} {\bibfnamefont
  {S.}~\bibnamefont {Sachdev}}, \bibinfo {author} {\bibfnamefont {M.~J.}\
  \bibnamefont {Lawler}}, \bibinfo {author} {\bibfnamefont {E.-A.}\
  \bibnamefont {Kim}}, \bibinfo {author} {\bibfnamefont {J.~C.}\ \bibnamefont
  {S{\'e}amus~Davis}}, \ and\ \bibinfo {author} {\bibfnamefont {M.~H.}\
  \bibnamefont {Hamidian}},\ }\href {\doibase 10.1126/science.aat1773}
  {\bibfield  {journal} {\bibinfo  {journal} {Science}\ }\textbf {\bibinfo
  {volume} {364}},\ \bibinfo {pages} {976} (\bibinfo {year}
  {2019})}\BibitemShut {NoStop}%
\bibitem [{\citenamefont {Berg}\ \emph {et~al.}(2007)\citenamefont {Berg},
  \citenamefont {Fradkin}, \citenamefont {Kim}, \citenamefont {Kivelson},
  \citenamefont {Oganesyan}, \citenamefont {Tranquada},\ and\ \citenamefont
  {Zhang}}]{Berg2007}%
  \BibitemOpen
  \bibfield  {author} {\bibinfo {author} {\bibfnamefont {E.}~\bibnamefont
  {Berg}}, \bibinfo {author} {\bibfnamefont {E.}~\bibnamefont {Fradkin}},
  \bibinfo {author} {\bibfnamefont {E.-A.}\ \bibnamefont {Kim}}, \bibinfo
  {author} {\bibfnamefont {S.~A.}\ \bibnamefont {Kivelson}}, \bibinfo {author}
  {\bibfnamefont {V.}~\bibnamefont {Oganesyan}}, \bibinfo {author}
  {\bibfnamefont {J.~M.}\ \bibnamefont {Tranquada}}, \ and\ \bibinfo {author}
  {\bibfnamefont {S.~C.}\ \bibnamefont {Zhang}},\ }\href {\doibase
  10.1103/PhysRevLett.99.127003} {\bibfield  {journal} {\bibinfo  {journal}
  {Phys. Rev. Lett.}\ }\textbf {\bibinfo {volume} {99}},\ \bibinfo {pages}
  {127003} (\bibinfo {year} {2007})}\BibitemShut {NoStop}%
\bibitem [{\citenamefont {Agterberg}\ and\ \citenamefont
  {Tsunetsugu}(2008)}]{Agterberg2008}%
  \BibitemOpen
  \bibfield  {author} {\bibinfo {author} {\bibfnamefont {D.~F.}\ \bibnamefont
  {Agterberg}}\ and\ \bibinfo {author} {\bibfnamefont {H.}~\bibnamefont
  {Tsunetsugu}},\ }\href {\doibase 10.1038/nphys999} {\bibfield  {journal}
  {\bibinfo  {journal} {Nature Physics}\ ,\ \bibinfo {pages} {639}} (\bibinfo
  {year} {2008})}\BibitemShut {NoStop}%
\bibitem [{\citenamefont {Berg}\ \emph {et~al.}(2010)\citenamefont {Berg},
  \citenamefont {Fradkin},\ and\ \citenamefont {Kivelson}}]{Berg2010}%
  \BibitemOpen
  \bibfield  {author} {\bibinfo {author} {\bibfnamefont {E.}~\bibnamefont
  {Berg}}, \bibinfo {author} {\bibfnamefont {E.}~\bibnamefont {Fradkin}}, \
  and\ \bibinfo {author} {\bibfnamefont {S.~A.}\ \bibnamefont {Kivelson}},\
  }\href {\doibase 10.1103/PhysRevLett.105.146403} {\bibfield  {journal}
  {\bibinfo  {journal} {Phys. Rev. Lett.}\ }\textbf {\bibinfo {volume} {105}},\
  \bibinfo {pages} {146403} (\bibinfo {year} {2010})}\BibitemShut {NoStop}%
\bibitem [{\citenamefont {Jaefari}\ and\ \citenamefont
  {Fradkin}(2012)}]{Fradkin2012}%
  \BibitemOpen
  \bibfield  {author} {\bibinfo {author} {\bibfnamefont {A.}~\bibnamefont
  {Jaefari}}\ and\ \bibinfo {author} {\bibfnamefont {E.}~\bibnamefont
  {Fradkin}},\ }\href {\doibase 10.1103/PhysRevB.85.035104} {\bibfield
  {journal} {\bibinfo  {journal} {Phys. Rev. B}\ }\textbf {\bibinfo {volume}
  {85}},\ \bibinfo {pages} {035104} (\bibinfo {year} {2012})}\BibitemShut
  {NoStop}%
\bibitem [{\citenamefont {Venderley}\ and\ \citenamefont
  {Kim}(2019)}]{Venderley2019}%
  \BibitemOpen
  \bibfield  {author} {\bibinfo {author} {\bibfnamefont {J.}~\bibnamefont
  {Venderley}}\ and\ \bibinfo {author} {\bibfnamefont {E.-A.}\ \bibnamefont
  {Kim}},\ }\href {\doibase 10.1103/PhysRevB.100.060506} {\bibfield  {journal}
  {\bibinfo  {journal} {Phys. Rev. B}\ }\textbf {\bibinfo {volume} {100}},\
  \bibinfo {pages} {060506} (\bibinfo {year} {2019})}\BibitemShut {NoStop}%
\bibitem [{\citenamefont {Xu}\ \emph {et~al.}(2019)\citenamefont {Xu},
  \citenamefont {Law},\ and\ \citenamefont {Lee}}]{Xu2019}%
  \BibitemOpen
  \bibfield  {author} {\bibinfo {author} {\bibfnamefont {X.~Y.}\ \bibnamefont
  {Xu}}, \bibinfo {author} {\bibfnamefont {K.~T.}\ \bibnamefont {Law}}, \ and\
  \bibinfo {author} {\bibfnamefont {P.~A.}\ \bibnamefont {Lee}},\ }\href
  {\doibase 10.1103/PhysRevLett.122.167001} {\bibfield  {journal} {\bibinfo
  {journal} {Phys. Rev. Lett.}\ }\textbf {\bibinfo {volume} {122}},\ \bibinfo
  {pages} {167001} (\bibinfo {year} {2019})}\BibitemShut {NoStop}%
\bibitem [{\citenamefont {Han}\ \emph {et~al.}(2020)\citenamefont {Han},
  \citenamefont {Kivelson},\ and\ \citenamefont {Yao}}]{Han2020}%
  \BibitemOpen
  \bibfield  {author} {\bibinfo {author} {\bibfnamefont {Z.}~\bibnamefont
  {Han}}, \bibinfo {author} {\bibfnamefont {S.~A.}\ \bibnamefont {Kivelson}}, \
  and\ \bibinfo {author} {\bibfnamefont {H.}~\bibnamefont {Yao}},\ }\href
  {\doibase 10.1103/PhysRevLett.125.167001} {\bibfield  {journal} {\bibinfo
  {journal} {Phys. Rev. Lett.}\ }\textbf {\bibinfo {volume} {125}},\ \bibinfo
  {pages} {167001} (\bibinfo {year} {2020})}\BibitemShut {NoStop}%
\bibitem [{\citenamefont {{Huang}}\ \emph {et~al.}(2021)\citenamefont
  {{Huang}}, \citenamefont {{Han}}, \citenamefont {{Kivelson}},\ and\
  \citenamefont {{Yao}}}]{Huang2021}%
  \BibitemOpen
  \bibfield  {author} {\bibinfo {author} {\bibfnamefont {K.~S.}\ \bibnamefont
  {{Huang}}}, \bibinfo {author} {\bibfnamefont {Z.}~\bibnamefont {{Han}}},
  \bibinfo {author} {\bibfnamefont {S.~A.}\ \bibnamefont {{Kivelson}}}, \ and\
  \bibinfo {author} {\bibfnamefont {H.}~\bibnamefont {{Yao}}},\ }\href@noop {}
  {\ ,\ \bibinfo {eid} {arXiv:2103.04984} (\bibinfo {year} {2021})}\BibitemShut
  {NoStop}%
\bibitem [{\citenamefont {{Peng}}\ \emph {et~al.}(2020)\citenamefont {{Peng}},
  \citenamefont {{Jiang}}, \citenamefont {{Devereaux}},\ and\ \citenamefont
  {{Jiang}}}]{Cheng2020}%
  \BibitemOpen
  \bibfield  {author} {\bibinfo {author} {\bibfnamefont {C.}~\bibnamefont
  {{Peng}}}, \bibinfo {author} {\bibfnamefont {Y.-F.}\ \bibnamefont {{Jiang}}},
  \bibinfo {author} {\bibfnamefont {T.~P.}\ \bibnamefont {{Devereaux}}}, \ and\
  \bibinfo {author} {\bibfnamefont {H.-C.}\ \bibnamefont {{Jiang}}},\
  }\href@noop {} {\ ,\ \bibinfo {eid} {arXiv:2008.03858} (\bibinfo {year}
  {2020})}\BibitemShut {NoStop}%
\bibitem [{\citenamefont {Dagotto}(1994)}]{Dagotto:1994cz}%
  \BibitemOpen
  \bibfield  {author} {\bibinfo {author} {\bibfnamefont {E.}~\bibnamefont
  {Dagotto}},\ }\href@noop {} {\bibfield  {journal} {\bibinfo  {journal} {Rev.
  Mod. Phys.}\ }\textbf {\bibinfo {volume} {66}},\ \bibinfo {pages} {763}
  (\bibinfo {year} {1994})}\BibitemShut {NoStop}%
\bibitem [{\citenamefont {Zhang}\ and\ \citenamefont
  {Rice}(1988)}]{zhang1988effective}%
  \BibitemOpen
  \bibfield  {author} {\bibinfo {author} {\bibfnamefont {F.}~\bibnamefont
  {Zhang}}\ and\ \bibinfo {author} {\bibfnamefont {T.}~\bibnamefont {Rice}},\
  }\href@noop {} {\bibfield  {journal} {\bibinfo  {journal} {Phys. Rev. B}\
  }\textbf {\bibinfo {volume} {37}},\ \bibinfo {pages} {3759} (\bibinfo {year}
  {1988})}\BibitemShut {NoStop}%
\bibitem [{\citenamefont {Lee}\ \emph {et~al.}(2006)\citenamefont {Lee},
  \citenamefont {Nagaosa},\ and\ \citenamefont {Wen}}]{Lee2006}%
  \BibitemOpen
  \bibfield  {author} {\bibinfo {author} {\bibfnamefont {P.~A.}\ \bibnamefont
  {Lee}}, \bibinfo {author} {\bibfnamefont {N.}~\bibnamefont {Nagaosa}}, \ and\
  \bibinfo {author} {\bibfnamefont {X.-G.}\ \bibnamefont {Wen}},\ }\href
  {\doibase 10.1103/RevModPhys.78.17} {\bibfield  {journal} {\bibinfo
  {journal} {Rev. Mod. Phys.}\ }\textbf {\bibinfo {volume} {78}},\ \bibinfo
  {pages} {17} (\bibinfo {year} {2006})}\BibitemShut {NoStop}%
\bibitem [{\citenamefont {Fradkin}\ \emph {et~al.}(2015)\citenamefont
  {Fradkin}, \citenamefont {Kivelson},\ and\ \citenamefont
  {Tranquada}}]{Fradkin2015}%
  \BibitemOpen
  \bibfield  {author} {\bibinfo {author} {\bibfnamefont {E.}~\bibnamefont
  {Fradkin}}, \bibinfo {author} {\bibfnamefont {S.~A.}\ \bibnamefont
  {Kivelson}}, \ and\ \bibinfo {author} {\bibfnamefont {J.~M.}\ \bibnamefont
  {Tranquada}},\ }\href@noop {} {\bibfield  {journal} {\bibinfo  {journal}
  {Rev. Mod. Phys.}\ }\textbf {\bibinfo {volume} {87}},\ \bibinfo {pages} {457}
  (\bibinfo {year} {2015})}\BibitemShut {NoStop}%
\bibitem [{\citenamefont {Anderson}(1973)}]{Anderson1973}%
  \BibitemOpen
  \bibfield  {author} {\bibinfo {author} {\bibfnamefont {P.}~\bibnamefont
  {Anderson}},\ }\href {\doibase https://doi.org/10.1016/0025-5408(73)90167-0}
  {\bibfield  {journal} {\bibinfo  {journal} {Materials Research Bulletin}\
  }\textbf {\bibinfo {volume} {8}},\ \bibinfo {pages} {153 } (\bibinfo {year}
  {1973})}\BibitemShut {NoStop}%
\bibitem [{\citenamefont {Balents}(2010)}]{Balents2010}%
  \BibitemOpen
  \bibfield  {author} {\bibinfo {author} {\bibfnamefont {L.}~\bibnamefont
  {Balents}},\ }\href {\doibase
  https://doi-org.stanford.idm.oclc.org/10.1038/nature08917} {\bibfield
  {journal} {\bibinfo  {journal} {Nature}\ }\textbf {\bibinfo {volume} {464}},\
  \bibinfo {pages} {199–208} (\bibinfo {year} {2010})}\BibitemShut {NoStop}%
\bibitem [{\citenamefont {Shimizu}\ \emph {et~al.}(2003)\citenamefont
  {Shimizu}, \citenamefont {Miyagawa}, \citenamefont {Kanoda}, \citenamefont
  {Maesato},\ and\ \citenamefont {Saito}}]{Shimizu2003}%
  \BibitemOpen
  \bibfield  {author} {\bibinfo {author} {\bibfnamefont {Y.}~\bibnamefont
  {Shimizu}}, \bibinfo {author} {\bibfnamefont {K.}~\bibnamefont {Miyagawa}},
  \bibinfo {author} {\bibfnamefont {K.}~\bibnamefont {Kanoda}}, \bibinfo
  {author} {\bibfnamefont {M.}~\bibnamefont {Maesato}}, \ and\ \bibinfo
  {author} {\bibfnamefont {G.}~\bibnamefont {Saito}},\ }\href {\doibase
  10.1103/PhysRevLett.91.107001} {\bibfield  {journal} {\bibinfo  {journal}
  {Phys. Rev. Lett.}\ }\textbf {\bibinfo {volume} {91}},\ \bibinfo {pages}
  {107001} (\bibinfo {year} {2003})}\BibitemShut {NoStop}%
\bibitem [{\citenamefont {Itou}\ \emph {et~al.}(2007)\citenamefont {Itou},
  \citenamefont {Oyamada}, \citenamefont {Maegawa}, \citenamefont {Tamura},\
  and\ \citenamefont {Kato}}]{Itou2007}%
  \BibitemOpen
  \bibfield  {author} {\bibinfo {author} {\bibfnamefont {T.}~\bibnamefont
  {Itou}}, \bibinfo {author} {\bibfnamefont {A.}~\bibnamefont {Oyamada}},
  \bibinfo {author} {\bibfnamefont {S.}~\bibnamefont {Maegawa}}, \bibinfo
  {author} {\bibfnamefont {M.}~\bibnamefont {Tamura}}, \ and\ \bibinfo {author}
  {\bibfnamefont {R.}~\bibnamefont {Kato}},\ }\href {\doibase
  10.1088/0953-8984/19/14/145247} {\bibfield  {journal} {\bibinfo  {journal}
  {Journal of Physics: Condensed Matter}\ }\textbf {\bibinfo {volume} {19}},\
  \bibinfo {pages} {145247} (\bibinfo {year} {2007})}\BibitemShut {NoStop}%
\bibitem [{\citenamefont {Itou}\ \emph {et~al.}(2008)\citenamefont {Itou},
  \citenamefont {Oyamada}, \citenamefont {Maegawa}, \citenamefont {Tamura},\
  and\ \citenamefont {Kato}}]{Itou2008}%
  \BibitemOpen
  \bibfield  {author} {\bibinfo {author} {\bibfnamefont {T.}~\bibnamefont
  {Itou}}, \bibinfo {author} {\bibfnamefont {A.}~\bibnamefont {Oyamada}},
  \bibinfo {author} {\bibfnamefont {S.}~\bibnamefont {Maegawa}}, \bibinfo
  {author} {\bibfnamefont {M.}~\bibnamefont {Tamura}}, \ and\ \bibinfo {author}
  {\bibfnamefont {R.}~\bibnamefont {Kato}},\ }\href {\doibase
  10.1103/PhysRevB.77.104413} {\bibfield  {journal} {\bibinfo  {journal} {Phys.
  Rev. B}\ }\textbf {\bibinfo {volume} {77}},\ \bibinfo {pages} {104413}
  (\bibinfo {year} {2008})}\BibitemShut {NoStop}%
\bibitem [{\citenamefont {Itou}\ \emph {et~al.}(2010)\citenamefont {Itou},
  \citenamefont {Oyamada}, \citenamefont {Maegawa},\ and\ \citenamefont
  {Kato}}]{Itou2010}%
  \BibitemOpen
  \bibfield  {author} {\bibinfo {author} {\bibfnamefont {T.}~\bibnamefont
  {Itou}}, \bibinfo {author} {\bibfnamefont {A.}~\bibnamefont {Oyamada}},
  \bibinfo {author} {\bibfnamefont {S.}~\bibnamefont {Maegawa}}, \ and\
  \bibinfo {author} {\bibfnamefont {R.}~\bibnamefont {Kato}},\ }\href {\doibase
  10.1038/nphys1715} {\bibfield  {journal} {\bibinfo  {journal} {Nature
  Physics}\ }\textbf {\bibinfo {volume} {6}},\ \bibinfo {pages} {673–676}
  (\bibinfo {year} {2010})}\BibitemShut {NoStop}%
\bibitem [{\citenamefont {Yamashita}\ \emph {et~al.}(2010)\citenamefont
  {Yamashita}, \citenamefont {Nakata}, \citenamefont {Senshu}, \citenamefont
  {Nagata}, \citenamefont {Yamamoto}, \citenamefont {Kato}, \citenamefont
  {Shibauchi},\ and\ \citenamefont {Matsuda}}]{Yamashita2010}%
  \BibitemOpen
  \bibfield  {author} {\bibinfo {author} {\bibfnamefont {M.}~\bibnamefont
  {Yamashita}}, \bibinfo {author} {\bibfnamefont {N.}~\bibnamefont {Nakata}},
  \bibinfo {author} {\bibfnamefont {Y.}~\bibnamefont {Senshu}}, \bibinfo
  {author} {\bibfnamefont {M.}~\bibnamefont {Nagata}}, \bibinfo {author}
  {\bibfnamefont {H.~M.}\ \bibnamefont {Yamamoto}}, \bibinfo {author}
  {\bibfnamefont {R.}~\bibnamefont {Kato}}, \bibinfo {author} {\bibfnamefont
  {T.}~\bibnamefont {Shibauchi}}, \ and\ \bibinfo {author} {\bibfnamefont
  {Y.}~\bibnamefont {Matsuda}},\ }\href {\doibase 10.1126/science.1188200}
  {\bibfield  {journal} {\bibinfo  {journal} {Science}\ }\textbf {\bibinfo
  {volume} {328}},\ \bibinfo {pages} {1246} (\bibinfo {year}
  {2010})}\BibitemShut {NoStop}%
\bibitem [{\citenamefont {Yamashita}\ \emph {et~al.}(2011)\citenamefont
  {Yamashita}, \citenamefont {Yamamoto}, \citenamefont {Nakazawa},
  \citenamefont {Tamura},\ and\ \citenamefont {Kato}}]{Yamashita2011}%
  \BibitemOpen
  \bibfield  {author} {\bibinfo {author} {\bibfnamefont {S.}~\bibnamefont
  {Yamashita}}, \bibinfo {author} {\bibfnamefont {T.}~\bibnamefont {Yamamoto}},
  \bibinfo {author} {\bibfnamefont {Y.}~\bibnamefont {Nakazawa}}, \bibinfo
  {author} {\bibfnamefont {M.}~\bibnamefont {Tamura}}, \ and\ \bibinfo {author}
  {\bibfnamefont {R.}~\bibnamefont {Kato}},\ }\href {\doibase
  10.1038/ncomms1274} {\bibfield  {journal} {\bibinfo  {journal} {Nature
  Communications}\ }\textbf {\bibinfo {volume} {2}},\ \bibinfo {pages} {275}
  (\bibinfo {year} {2011})}\BibitemShut {NoStop}%
\bibitem [{\citenamefont {Senthil}(2008)}]{Senthil2008}%
  \BibitemOpen
  \bibfield  {author} {\bibinfo {author} {\bibfnamefont {T.}~\bibnamefont
  {Senthil}},\ }\href {\doibase 10.1103/PhysRevB.78.045109} {\bibfield
  {journal} {\bibinfo  {journal} {Phys. Rev. B}\ }\textbf {\bibinfo {volume}
  {78}},\ \bibinfo {pages} {045109} (\bibinfo {year} {2008})}\BibitemShut
  {NoStop}%
\bibitem [{\citenamefont {Misguich}\ \emph {et~al.}(1999)\citenamefont
  {Misguich}, \citenamefont {Lhuillier}, \citenamefont {Bernu},\ and\
  \citenamefont {Waldtmann}}]{Misguich1999}%
  \BibitemOpen
  \bibfield  {author} {\bibinfo {author} {\bibfnamefont {G.}~\bibnamefont
  {Misguich}}, \bibinfo {author} {\bibfnamefont {C.}~\bibnamefont {Lhuillier}},
  \bibinfo {author} {\bibfnamefont {B.}~\bibnamefont {Bernu}}, \ and\ \bibinfo
  {author} {\bibfnamefont {C.}~\bibnamefont {Waldtmann}},\ }\href {\doibase
  10.1103/PhysRevB.60.1064} {\bibfield  {journal} {\bibinfo  {journal} {Phys.
  Rev. B}\ }\textbf {\bibinfo {volume} {60}},\ \bibinfo {pages} {1064}
  (\bibinfo {year} {1999})}\BibitemShut {NoStop}%
\bibitem [{\citenamefont {LiMing}\ \emph {et~al.}(2000)\citenamefont {LiMing},
  \citenamefont {Misguich}, \citenamefont {Sindzingre},\ and\ \citenamefont
  {Lhuillier}}]{LiMingW2000}%
  \BibitemOpen
  \bibfield  {author} {\bibinfo {author} {\bibfnamefont {W.}~\bibnamefont
  {LiMing}}, \bibinfo {author} {\bibfnamefont {G.}~\bibnamefont {Misguich}},
  \bibinfo {author} {\bibfnamefont {P.}~\bibnamefont {Sindzingre}}, \ and\
  \bibinfo {author} {\bibfnamefont {C.}~\bibnamefont {Lhuillier}},\ }\href
  {\doibase 10.1103/PhysRevB.62.6372} {\bibfield  {journal} {\bibinfo
  {journal} {Phys. Rev. B}\ }\textbf {\bibinfo {volume} {62}},\ \bibinfo
  {pages} {6372} (\bibinfo {year} {2000})}\BibitemShut {NoStop}%
\bibitem [{\citenamefont {Mishmash}\ \emph {et~al.}(2013)\citenamefont
  {Mishmash}, \citenamefont {Garrison}, \citenamefont {Bieri},\ and\
  \citenamefont {Xu}}]{Mishmash2013}%
  \BibitemOpen
  \bibfield  {author} {\bibinfo {author} {\bibfnamefont {R.~V.}\ \bibnamefont
  {Mishmash}}, \bibinfo {author} {\bibfnamefont {J.~R.}\ \bibnamefont
  {Garrison}}, \bibinfo {author} {\bibfnamefont {S.}~\bibnamefont {Bieri}}, \
  and\ \bibinfo {author} {\bibfnamefont {C.}~\bibnamefont {Xu}},\ }\href
  {\doibase 10.1103/PhysRevLett.111.157203} {\bibfield  {journal} {\bibinfo
  {journal} {Phys. Rev. Lett.}\ }\textbf {\bibinfo {volume} {111}},\ \bibinfo
  {pages} {157203} (\bibinfo {year} {2013})}\BibitemShut {NoStop}%
\bibitem [{\citenamefont {Kyung}\ and\ \citenamefont
  {Tremblay}(2006)}]{Kyung2006}%
  \BibitemOpen
  \bibfield  {author} {\bibinfo {author} {\bibfnamefont {B.}~\bibnamefont
  {Kyung}}\ and\ \bibinfo {author} {\bibfnamefont {A.-M.~S.}\ \bibnamefont
  {Tremblay}},\ }\href {\doibase 10.1103/PhysRevLett.97.046402} {\bibfield
  {journal} {\bibinfo  {journal} {Phys. Rev. Lett.}\ }\textbf {\bibinfo
  {volume} {97}},\ \bibinfo {pages} {046402} (\bibinfo {year}
  {2006})}\BibitemShut {NoStop}%
\bibitem [{\citenamefont {Clay}\ \emph {et~al.}(2008)\citenamefont {Clay},
  \citenamefont {Li},\ and\ \citenamefont {Mazumdar}}]{ClayRT2008}%
  \BibitemOpen
  \bibfield  {author} {\bibinfo {author} {\bibfnamefont {R.~T.}\ \bibnamefont
  {Clay}}, \bibinfo {author} {\bibfnamefont {H.}~\bibnamefont {Li}}, \ and\
  \bibinfo {author} {\bibfnamefont {S.}~\bibnamefont {Mazumdar}},\ }\href
  {\doibase 10.1103/PhysRevLett.101.166403} {\bibfield  {journal} {\bibinfo
  {journal} {Phys. Rev. Lett.}\ }\textbf {\bibinfo {volume} {101}},\ \bibinfo
  {pages} {166403} (\bibinfo {year} {2008})}\BibitemShut {NoStop}%
\bibitem [{\citenamefont {Morita}\ \emph {et~al.}(2002)\citenamefont {Morita},
  \citenamefont {Watanabe},\ and\ \citenamefont {Imada}}]{Morita2002}%
  \BibitemOpen
  \bibfield  {author} {\bibinfo {author} {\bibfnamefont {H.}~\bibnamefont
  {Morita}}, \bibinfo {author} {\bibfnamefont {S.}~\bibnamefont {Watanabe}}, \
  and\ \bibinfo {author} {\bibfnamefont {M.}~\bibnamefont {Imada}},\ }\href
  {\doibase 10.1143/JPSJ.71.2109} {\bibfield  {journal} {\bibinfo  {journal}
  {Journal of the Physical Society of Japan}\ }\textbf {\bibinfo {volume}
  {71}},\ \bibinfo {pages} {2109} (\bibinfo {year} {2002})}\BibitemShut
  {NoStop}%
\bibitem [{\citenamefont {Koretsune}\ \emph {et~al.}(2007)\citenamefont
  {Koretsune}, \citenamefont {Motome},\ and\ \citenamefont
  {Furusaki}}]{Koretsune2007}%
  \BibitemOpen
  \bibfield  {author} {\bibinfo {author} {\bibfnamefont {T.}~\bibnamefont
  {Koretsune}}, \bibinfo {author} {\bibfnamefont {Y.}~\bibnamefont {Motome}}, \
  and\ \bibinfo {author} {\bibfnamefont {A.}~\bibnamefont {Furusaki}},\ }\href
  {\doibase 10.1143/JPSJ.76.074719} {\bibfield  {journal} {\bibinfo  {journal}
  {Journal of the Physical Society of Japan}\ }\textbf {\bibinfo {volume}
  {76}},\ \bibinfo {pages} {074719} (\bibinfo {year} {2007})}\BibitemShut
  {NoStop}%
\bibitem [{\citenamefont {Motrunich}(2005)}]{Motrunich2005}%
  \BibitemOpen
  \bibfield  {author} {\bibinfo {author} {\bibfnamefont {O.~I.}\ \bibnamefont
  {Motrunich}},\ }\href {\doibase 10.1103/PhysRevB.72.045105} {\bibfield
  {journal} {\bibinfo  {journal} {Phys. Rev. B}\ }\textbf {\bibinfo {volume}
  {72}},\ \bibinfo {pages} {045105} (\bibinfo {year} {2005})}\BibitemShut
  {NoStop}%
\bibitem [{\citenamefont {Yang}\ \emph {et~al.}(2010)\citenamefont {Yang},
  \citenamefont {L\"auchli}, \citenamefont {Mila},\ and\ \citenamefont
  {Schmidt}}]{Yang2010}%
  \BibitemOpen
  \bibfield  {author} {\bibinfo {author} {\bibfnamefont {H.-Y.}\ \bibnamefont
  {Yang}}, \bibinfo {author} {\bibfnamefont {A.~M.}\ \bibnamefont {L\"auchli}},
  \bibinfo {author} {\bibfnamefont {F.}~\bibnamefont {Mila}}, \ and\ \bibinfo
  {author} {\bibfnamefont {K.~P.}\ \bibnamefont {Schmidt}},\ }\href {\doibase
  10.1103/PhysRevLett.105.267204} {\bibfield  {journal} {\bibinfo  {journal}
  {Phys. Rev. Lett.}\ }\textbf {\bibinfo {volume} {105}},\ \bibinfo {pages}
  {267204} (\bibinfo {year} {2010})}\BibitemShut {NoStop}%
\bibitem [{\citenamefont {Sheng}\ \emph {et~al.}(2009)\citenamefont {Sheng},
  \citenamefont {Motrunich},\ and\ \citenamefont {Fisher}}]{ShengDN2009}%
  \BibitemOpen
  \bibfield  {author} {\bibinfo {author} {\bibfnamefont {D.~N.}\ \bibnamefont
  {Sheng}}, \bibinfo {author} {\bibfnamefont {O.~I.}\ \bibnamefont
  {Motrunich}}, \ and\ \bibinfo {author} {\bibfnamefont {M.~P.~A.}\
  \bibnamefont {Fisher}},\ }\href {\doibase 10.1103/PhysRevB.79.205112}
  {\bibfield  {journal} {\bibinfo  {journal} {Phys. Rev. B}\ }\textbf {\bibinfo
  {volume} {79}},\ \bibinfo {pages} {205112} (\bibinfo {year}
  {2009})}\BibitemShut {NoStop}%
\bibitem [{\citenamefont {Block}\ \emph {et~al.}(2011)\citenamefont {Block},
  \citenamefont {Sheng}, \citenamefont {Motrunich},\ and\ \citenamefont
  {Fisher}}]{Block2011}%
  \BibitemOpen
  \bibfield  {author} {\bibinfo {author} {\bibfnamefont {M.~S.}\ \bibnamefont
  {Block}}, \bibinfo {author} {\bibfnamefont {D.~N.}\ \bibnamefont {Sheng}},
  \bibinfo {author} {\bibfnamefont {O.~I.}\ \bibnamefont {Motrunich}}, \ and\
  \bibinfo {author} {\bibfnamefont {M.~P.~A.}\ \bibnamefont {Fisher}},\ }\href
  {\doibase 10.1103/PhysRevLett.106.157202} {\bibfield  {journal} {\bibinfo
  {journal} {Phys. Rev. Lett.}\ }\textbf {\bibinfo {volume} {106}},\ \bibinfo
  {pages} {157202} (\bibinfo {year} {2011})}\BibitemShut {NoStop}%
\bibitem [{\citenamefont {Hu}\ \emph {et~al.}(2015)\citenamefont {Hu},
  \citenamefont {Gong}, \citenamefont {Zhu},\ and\ \citenamefont
  {Sheng}}]{Hu2015}%
  \BibitemOpen
  \bibfield  {author} {\bibinfo {author} {\bibfnamefont {W.-J.}\ \bibnamefont
  {Hu}}, \bibinfo {author} {\bibfnamefont {S.-S.}\ \bibnamefont {Gong}},
  \bibinfo {author} {\bibfnamefont {W.}~\bibnamefont {Zhu}}, \ and\ \bibinfo
  {author} {\bibfnamefont {D.~N.}\ \bibnamefont {Sheng}},\ }\href {\doibase
  10.1103/PhysRevB.92.140403} {\bibfield  {journal} {\bibinfo  {journal} {Phys.
  Rev. B}\ }\textbf {\bibinfo {volume} {92}},\ \bibinfo {pages} {140403}
  (\bibinfo {year} {2015})}\BibitemShut {NoStop}%
\bibitem [{\citenamefont {Qi}\ and\ \citenamefont {Sachdev}(2008)}]{QiY2008}%
  \BibitemOpen
  \bibfield  {author} {\bibinfo {author} {\bibfnamefont {Y.}~\bibnamefont
  {Qi}}\ and\ \bibinfo {author} {\bibfnamefont {S.}~\bibnamefont {Sachdev}},\
  }\href {\doibase 10.1103/PhysRevB.77.165112} {\bibfield  {journal} {\bibinfo
  {journal} {Phys. Rev. B}\ }\textbf {\bibinfo {volume} {77}},\ \bibinfo
  {pages} {165112} (\bibinfo {year} {2008})}\BibitemShut {NoStop}%
\bibitem [{\citenamefont {Shirakawa}\ \emph {et~al.}(2017)\citenamefont
  {Shirakawa}, \citenamefont {Tohyama}, \citenamefont {Kokalj}, \citenamefont
  {Sota},\ and\ \citenamefont {Yunoki}}]{Shirakawa2017}%
  \BibitemOpen
  \bibfield  {author} {\bibinfo {author} {\bibfnamefont {T.}~\bibnamefont
  {Shirakawa}}, \bibinfo {author} {\bibfnamefont {T.}~\bibnamefont {Tohyama}},
  \bibinfo {author} {\bibfnamefont {J.}~\bibnamefont {Kokalj}}, \bibinfo
  {author} {\bibfnamefont {S.}~\bibnamefont {Sota}}, \ and\ \bibinfo {author}
  {\bibfnamefont {S.}~\bibnamefont {Yunoki}},\ }\href {\doibase
  10.1103/PhysRevB.96.205130} {\bibfield  {journal} {\bibinfo  {journal} {Phys.
  Rev. B}\ }\textbf {\bibinfo {volume} {96}},\ \bibinfo {pages} {205130}
  (\bibinfo {year} {2017})}\BibitemShut {NoStop}%
\bibitem [{\citenamefont {Sahebsara}\ and\ \citenamefont
  {S\'en\'echal}(2008)}]{Sahebsara2008}%
  \BibitemOpen
  \bibfield  {author} {\bibinfo {author} {\bibfnamefont {P.}~\bibnamefont
  {Sahebsara}}\ and\ \bibinfo {author} {\bibfnamefont {D.}~\bibnamefont
  {S\'en\'echal}},\ }\href {\doibase 10.1103/PhysRevLett.100.136402} {\bibfield
   {journal} {\bibinfo  {journal} {Phys. Rev. Lett.}\ }\textbf {\bibinfo
  {volume} {100}},\ \bibinfo {pages} {136402} (\bibinfo {year}
  {2008})}\BibitemShut {NoStop}%
\bibitem [{\citenamefont {Laubach}\ \emph {et~al.}(2015)\citenamefont
  {Laubach}, \citenamefont {Thomale}, \citenamefont {Platt}, \citenamefont
  {Hanke},\ and\ \citenamefont {Li}}]{Laubach2015}%
  \BibitemOpen
  \bibfield  {author} {\bibinfo {author} {\bibfnamefont {M.}~\bibnamefont
  {Laubach}}, \bibinfo {author} {\bibfnamefont {R.}~\bibnamefont {Thomale}},
  \bibinfo {author} {\bibfnamefont {C.}~\bibnamefont {Platt}}, \bibinfo
  {author} {\bibfnamefont {W.}~\bibnamefont {Hanke}}, \ and\ \bibinfo {author}
  {\bibfnamefont {G.}~\bibnamefont {Li}},\ }\href {\doibase
  10.1103/PhysRevB.91.245125} {\bibfield  {journal} {\bibinfo  {journal} {Phys.
  Rev. B}\ }\textbf {\bibinfo {volume} {91}},\ \bibinfo {pages} {245125}
  (\bibinfo {year} {2015})}\BibitemShut {NoStop}%
\bibitem [{\citenamefont {Yoshioka}\ \emph {et~al.}(2009)\citenamefont
  {Yoshioka}, \citenamefont {Koga},\ and\ \citenamefont
  {Kawakami}}]{Yoshioka2009}%
  \BibitemOpen
  \bibfield  {author} {\bibinfo {author} {\bibfnamefont {T.}~\bibnamefont
  {Yoshioka}}, \bibinfo {author} {\bibfnamefont {A.}~\bibnamefont {Koga}}, \
  and\ \bibinfo {author} {\bibfnamefont {N.}~\bibnamefont {Kawakami}},\ }\href
  {\doibase 10.1103/PhysRevLett.103.036401} {\bibfield  {journal} {\bibinfo
  {journal} {Phys. Rev. Lett.}\ }\textbf {\bibinfo {volume} {103}},\ \bibinfo
  {pages} {036401} (\bibinfo {year} {2009})}\BibitemShut {NoStop}%
\bibitem [{\citenamefont {Mizusaki}\ and\ \citenamefont
  {Imada}(2006)}]{Mizusaki2006}%
  \BibitemOpen
  \bibfield  {author} {\bibinfo {author} {\bibfnamefont {T.}~\bibnamefont
  {Mizusaki}}\ and\ \bibinfo {author} {\bibfnamefont {M.}~\bibnamefont
  {Imada}},\ }\href {\doibase 10.1103/PhysRevB.74.014421} {\bibfield  {journal}
  {\bibinfo  {journal} {Phys. Rev. B}\ }\textbf {\bibinfo {volume} {74}},\
  \bibinfo {pages} {014421} (\bibinfo {year} {2006})}\BibitemShut {NoStop}%
\bibitem [{\citenamefont {Szasz}\ \emph {et~al.}(2020)\citenamefont {Szasz},
  \citenamefont {Motruk}, \citenamefont {Zaletel},\ and\ \citenamefont
  {Moore}}]{Szasz2020}%
  \BibitemOpen
  \bibfield  {author} {\bibinfo {author} {\bibfnamefont {A.}~\bibnamefont
  {Szasz}}, \bibinfo {author} {\bibfnamefont {J.}~\bibnamefont {Motruk}},
  \bibinfo {author} {\bibfnamefont {M.~P.}\ \bibnamefont {Zaletel}}, \ and\
  \bibinfo {author} {\bibfnamefont {J.~E.}\ \bibnamefont {Moore}},\ }\href
  {\doibase 10.1103/PhysRevX.10.021042} {\bibfield  {journal} {\bibinfo
  {journal} {Phys. Rev. X}\ }\textbf {\bibinfo {volume} {10}},\ \bibinfo
  {pages} {021042} (\bibinfo {year} {2020})}\BibitemShut {NoStop}%
\bibitem [{\citenamefont {Mishmash}\ \emph {et~al.}(2015)\citenamefont
  {Mishmash}, \citenamefont {Gonz\'alez}, \citenamefont {Melko}, \citenamefont
  {Motrunich},\ and\ \citenamefont {Fisher}}]{Mishmash2015}%
  \BibitemOpen
  \bibfield  {author} {\bibinfo {author} {\bibfnamefont {R.~V.}\ \bibnamefont
  {Mishmash}}, \bibinfo {author} {\bibfnamefont {I.}~\bibnamefont
  {Gonz\'alez}}, \bibinfo {author} {\bibfnamefont {R.~G.}\ \bibnamefont
  {Melko}}, \bibinfo {author} {\bibfnamefont {O.~I.}\ \bibnamefont
  {Motrunich}}, \ and\ \bibinfo {author} {\bibfnamefont {M.~P.~A.}\
  \bibnamefont {Fisher}},\ }\href {\doibase 10.1103/PhysRevB.91.235140}
  {\bibfield  {journal} {\bibinfo  {journal} {Phys. Rev. B}\ }\textbf {\bibinfo
  {volume} {91}},\ \bibinfo {pages} {235140} (\bibinfo {year}
  {2015})}\BibitemShut {NoStop}%
\bibitem [{\citenamefont {White}(1992)}]{White1992}%
  \BibitemOpen
  \bibfield  {author} {\bibinfo {author} {\bibfnamefont {S.~R.}\ \bibnamefont
  {White}},\ }\href {\doibase 10.1103/PhysRevLett.69.2863} {\bibfield
  {journal} {\bibinfo  {journal} {Phys. Rev. Lett.}\ }\textbf {\bibinfo
  {volume} {69}},\ \bibinfo {pages} {2863} (\bibinfo {year}
  {1992})}\BibitemShut {NoStop}%
\bibitem [{\citenamefont {{Chen}}\ \emph {et~al.}(2021)\citenamefont {{Chen}},
  \citenamefont {{Chen}}, \citenamefont {{Gong}}, \citenamefont {{Sheng}},
  \citenamefont {{Li}},\ and\ \citenamefont {{Weichselbaum}}}]{Chen2021}%
  \BibitemOpen
  \bibfield  {author} {\bibinfo {author} {\bibfnamefont {B.-B.}\ \bibnamefont
  {{Chen}}}, \bibinfo {author} {\bibfnamefont {Z.}~\bibnamefont {{Chen}}},
  \bibinfo {author} {\bibfnamefont {S.-S.}\ \bibnamefont {{Gong}}}, \bibinfo
  {author} {\bibfnamefont {D.~N.}\ \bibnamefont {{Sheng}}}, \bibinfo {author}
  {\bibfnamefont {W.}~\bibnamefont {{Li}}}, \ and\ \bibinfo {author}
  {\bibfnamefont {A.}~\bibnamefont {{Weichselbaum}}},\ }\href@noop {} {\ ,\
  \bibinfo {eid} {arXiv:2102.05560} (\bibinfo {year} {2021})}\BibitemShut
  {NoStop}%
\bibitem [{\citenamefont {ANDERSON}(1987)}]{Anderson1987}%
  \BibitemOpen
  \bibfield  {author} {\bibinfo {author} {\bibfnamefont {P.~W.}\ \bibnamefont
  {ANDERSON}},\ }\href {\doibase 10.1126/science.235.4793.1196} {\bibfield
  {journal} {\bibinfo  {journal} {Science}\ }\textbf {\bibinfo {volume}
  {235}},\ \bibinfo {pages} {1196} (\bibinfo {year} {1987})}\BibitemShut
  {NoStop}%
\bibitem [{\citenamefont {Kivelson}\ \emph {et~al.}(1987)\citenamefont
  {Kivelson}, \citenamefont {Rokhsar},\ and\ \citenamefont
  {Sethna}}]{Kivelson1987}%
  \BibitemOpen
  \bibfield  {author} {\bibinfo {author} {\bibfnamefont {S.~A.}\ \bibnamefont
  {Kivelson}}, \bibinfo {author} {\bibfnamefont {D.~S.}\ \bibnamefont
  {Rokhsar}}, \ and\ \bibinfo {author} {\bibfnamefont {J.~P.}\ \bibnamefont
  {Sethna}},\ }\href {\doibase 10.1103/PhysRevB.35.8865} {\bibfield  {journal}
  {\bibinfo  {journal} {Phys. Rev. B}\ }\textbf {\bibinfo {volume} {35}},\
  \bibinfo {pages} {8865} (\bibinfo {year} {1987})}\BibitemShut {NoStop}%
\bibitem [{\citenamefont {Rokhsar}\ and\ \citenamefont
  {Kivelson}(1988)}]{Rokhsar1988}%
  \BibitemOpen
  \bibfield  {author} {\bibinfo {author} {\bibfnamefont {D.~S.}\ \bibnamefont
  {Rokhsar}}\ and\ \bibinfo {author} {\bibfnamefont {S.~A.}\ \bibnamefont
  {Kivelson}},\ }\href@noop {} {\bibfield  {journal} {\bibinfo  {journal}
  {Phys. Rev. Lett.}\ }\textbf {\bibinfo {volume} {61}},\ \bibinfo {pages}
  {2376} (\bibinfo {year} {1988})}\BibitemShut {NoStop}%
\bibitem [{\citenamefont {Laughlin}(1988)}]{Laughlin1988}%
  \BibitemOpen
  \bibfield  {author} {\bibinfo {author} {\bibfnamefont {R.~B.}\ \bibnamefont
  {Laughlin}},\ }\href@noop {} {\bibfield  {journal} {\bibinfo  {journal}
  {Science}\ }\textbf {\bibinfo {volume} {242}},\ \bibinfo {pages} {525}
  (\bibinfo {year} {1988})}\BibitemShut {NoStop}%
\bibitem [{\citenamefont {Wen}\ \emph {et~al.}(1989)\citenamefont {Wen},
  \citenamefont {Wilczek},\ and\ \citenamefont {Zee}}]{Wen1989}%
  \BibitemOpen
  \bibfield  {author} {\bibinfo {author} {\bibfnamefont {X.~G.}\ \bibnamefont
  {Wen}}, \bibinfo {author} {\bibfnamefont {F.}~\bibnamefont {Wilczek}}, \ and\
  \bibinfo {author} {\bibfnamefont {A.}~\bibnamefont {Zee}},\ }\href@noop {}
  {\bibfield  {journal} {\bibinfo  {journal} {Phys. Rev. B}\ }\textbf {\bibinfo
  {volume} {39}},\ \bibinfo {pages} {11413} (\bibinfo {year}
  {1989})}\BibitemShut {NoStop}%
\bibitem [{\citenamefont {Broholm}\ \emph {et~al.}(2020)\citenamefont
  {Broholm}, \citenamefont {Cava}, \citenamefont {Kivelson}, \citenamefont
  {Nocera}, \citenamefont {Norman},\ and\ \citenamefont
  {Senthil}}]{Broholm2019}%
  \BibitemOpen
  \bibfield  {author} {\bibinfo {author} {\bibfnamefont {C.}~\bibnamefont
  {Broholm}}, \bibinfo {author} {\bibfnamefont {R.~J.}\ \bibnamefont {Cava}},
  \bibinfo {author} {\bibfnamefont {S.~A.}\ \bibnamefont {Kivelson}}, \bibinfo
  {author} {\bibfnamefont {D.~G.}\ \bibnamefont {Nocera}}, \bibinfo {author}
  {\bibfnamefont {M.~R.}\ \bibnamefont {Norman}}, \ and\ \bibinfo {author}
  {\bibfnamefont {T.}~\bibnamefont {Senthil}},\ }\href@noop {} {\bibfield
  {journal} {\bibinfo  {journal} {Science}\ }\textbf {\bibinfo {volume}
  {367}},\ \bibinfo {pages} {eaay0668} (\bibinfo {year} {2020})}\BibitemShut
  {NoStop}%
\bibitem [{\citenamefont {Jiang}(2019)}]{Jiang2019tqsl}%
  \BibitemOpen
  \bibfield  {author} {\bibinfo {author} {\bibfnamefont {H.-C.}\ \bibnamefont
  {Jiang}},\ }\href@noop {} {\  (\bibinfo {year} {2019})},\ \Eprint
  {http://arxiv.org/abs/1912.06624} {arXiv:1912.06624} \BibitemShut {NoStop}%
\bibitem [{\citenamefont {Jiang}\ and\ \citenamefont
  {Jiang}(2020)}]{Jiang2020tcsl}%
  \BibitemOpen
  \bibfield  {author} {\bibinfo {author} {\bibfnamefont {Y.-F.}\ \bibnamefont
  {Jiang}}\ and\ \bibinfo {author} {\bibfnamefont {H.-C.}\ \bibnamefont
  {Jiang}},\ }\href {\doibase 10.1103/PhysRevLett.125.157002} {\bibfield
  {journal} {\bibinfo  {journal} {Phys. Rev. Lett.}\ }\textbf {\bibinfo
  {volume} {125}},\ \bibinfo {pages} {157002} (\bibinfo {year}
  {2020})}\BibitemShut {NoStop}%
\bibitem [{\citenamefont {Raghu}\ \emph {et~al.}(2010)\citenamefont {Raghu},
  \citenamefont {Kivelson},\ and\ \citenamefont {Scalapino}}]{Raghu2010}%
  \BibitemOpen
  \bibfield  {author} {\bibinfo {author} {\bibfnamefont {S.}~\bibnamefont
  {Raghu}}, \bibinfo {author} {\bibfnamefont {S.~A.}\ \bibnamefont {Kivelson}},
  \ and\ \bibinfo {author} {\bibfnamefont {D.~J.}\ \bibnamefont {Scalapino}},\
  }\href {\doibase 10.1103/PhysRevB.81.224505} {\bibfield  {journal} {\bibinfo
  {journal} {Phys. Rev. B}\ }\textbf {\bibinfo {volume} {81}},\ \bibinfo
  {pages} {224505} (\bibinfo {year} {2010})}\BibitemShut {NoStop}%
\bibitem [{\citenamefont {Chen}\ \emph
  {et~al.}(2013{\natexlab{a}})\citenamefont {Chen}, \citenamefont {Meng},
  \citenamefont {Yu}, \citenamefont {Yang}, \citenamefont {Jarrell},\ and\
  \citenamefont {Moreno}}]{ChenKS2013}%
  \BibitemOpen
  \bibfield  {author} {\bibinfo {author} {\bibfnamefont {K.~S.}\ \bibnamefont
  {Chen}}, \bibinfo {author} {\bibfnamefont {Z.~Y.}\ \bibnamefont {Meng}},
  \bibinfo {author} {\bibfnamefont {U.}~\bibnamefont {Yu}}, \bibinfo {author}
  {\bibfnamefont {S.}~\bibnamefont {Yang}}, \bibinfo {author} {\bibfnamefont
  {M.}~\bibnamefont {Jarrell}}, \ and\ \bibinfo {author} {\bibfnamefont
  {J.}~\bibnamefont {Moreno}},\ }\href {\doibase 10.1103/PhysRevB.88.041103}
  {\bibfield  {journal} {\bibinfo  {journal} {Phys. Rev. B}\ }\textbf {\bibinfo
  {volume} {88}},\ \bibinfo {pages} {041103} (\bibinfo {year}
  {2013}{\natexlab{a}})}\BibitemShut {NoStop}%
\bibitem [{\citenamefont {Guo}\ \emph {et~al.}(2018)\citenamefont {Guo},
  \citenamefont {Zhu}, \citenamefont {Feng},\ and\ \citenamefont
  {Scalettar}}]{GuoHM2018}%
  \BibitemOpen
  \bibfield  {author} {\bibinfo {author} {\bibfnamefont {H.}~\bibnamefont
  {Guo}}, \bibinfo {author} {\bibfnamefont {X.}~\bibnamefont {Zhu}}, \bibinfo
  {author} {\bibfnamefont {S.}~\bibnamefont {Feng}}, \ and\ \bibinfo {author}
  {\bibfnamefont {R.~T.}\ \bibnamefont {Scalettar}},\ }\href {\doibase
  10.1103/PhysRevB.97.235453} {\bibfield  {journal} {\bibinfo  {journal} {Phys.
  Rev. B}\ }\textbf {\bibinfo {volume} {97}},\ \bibinfo {pages} {235453}
  (\bibinfo {year} {2018})}\BibitemShut {NoStop}%
\bibitem [{\citenamefont {Zhu}\ \emph {et~al.}(2020)\citenamefont {Zhu},
  \citenamefont {Sheng},\ and\ \citenamefont {Vishwanath}}]{Zhu2020}%
  \BibitemOpen
  \bibfield  {author} {\bibinfo {author} {\bibfnamefont {Z.}~\bibnamefont
  {Zhu}}, \bibinfo {author} {\bibfnamefont {D.~N.}\ \bibnamefont {Sheng}}, \
  and\ \bibinfo {author} {\bibfnamefont {A.}~\bibnamefont {Vishwanath}},\
  }\href@noop {} {\  (\bibinfo {year} {2020})},\ \Eprint
  {http://arxiv.org/abs/2007.11963} {arXiv:2007.11963} \BibitemShut {NoStop}%
\bibitem [{\citenamefont {Chen}\ \emph
  {et~al.}(2013{\natexlab{b}})\citenamefont {Chen}, \citenamefont {Ju},
  \citenamefont {Jiang}, \citenamefont {Starykh},\ and\ \citenamefont
  {Balents}}]{Ru2013}%
  \BibitemOpen
  \bibfield  {author} {\bibinfo {author} {\bibfnamefont {R.}~\bibnamefont
  {Chen}}, \bibinfo {author} {\bibfnamefont {H.}~\bibnamefont {Ju}}, \bibinfo
  {author} {\bibfnamefont {H.-C.}\ \bibnamefont {Jiang}}, \bibinfo {author}
  {\bibfnamefont {O.~A.}\ \bibnamefont {Starykh}}, \ and\ \bibinfo {author}
  {\bibfnamefont {L.}~\bibnamefont {Balents}},\ }\href {\doibase
  10.1103/PhysRevB.87.165123} {\bibfield  {journal} {\bibinfo  {journal} {Phys.
  Rev. B}\ }\textbf {\bibinfo {volume} {87}},\ \bibinfo {pages} {165123}
  (\bibinfo {year} {2013}{\natexlab{b}})}\BibitemShut {NoStop}%
\bibitem [{\citenamefont {Calabrese}\ and\ \citenamefont
  {Cardy}(2004)}]{Calabrese2004}%
  \BibitemOpen
  \bibfield  {author} {\bibinfo {author} {\bibfnamefont {P.}~\bibnamefont
  {Calabrese}}\ and\ \bibinfo {author} {\bibfnamefont {J.}~\bibnamefont
  {Cardy}},\ }\href@noop {} {\bibfield  {journal} {\bibinfo  {journal} {J.
  Stat. Mech. Theory Exp.}\ }\textbf {\bibinfo {volume} {2004}} (\bibinfo
  {year} {2004})}\BibitemShut {NoStop}%
\bibitem [{\citenamefont {Fagotti}\ and\ \citenamefont
  {Calabrese}(2011)}]{Fagotti2011}%
  \BibitemOpen
  \bibfield  {author} {\bibinfo {author} {\bibfnamefont {M.}~\bibnamefont
  {Fagotti}}\ and\ \bibinfo {author} {\bibfnamefont {P.}~\bibnamefont
  {Calabrese}},\ }\href@noop {} {\bibfield  {journal} {\bibinfo  {journal} {J.
  Stat. Mech. Theory Exp.}\ }\textbf {\bibinfo {volume} {2011}} (\bibinfo
  {year} {2011})}\BibitemShut {NoStop}%
\bibitem [{\citenamefont {White}\ \emph {et~al.}(2002)\citenamefont {White},
  \citenamefont {Affleck},\ and\ \citenamefont {Scalapino}}]{White2002}%
  \BibitemOpen
  \bibfield  {author} {\bibinfo {author} {\bibfnamefont {S.~R.}\ \bibnamefont
  {White}}, \bibinfo {author} {\bibfnamefont {I.}~\bibnamefont {Affleck}}, \
  and\ \bibinfo {author} {\bibfnamefont {D.~J.}\ \bibnamefont {Scalapino}},\
  }\href {\doibase 10.1103/PhysRevB.65.165122} {\bibfield  {journal} {\bibinfo
  {journal} {Phys. Rev. B}\ }\textbf {\bibinfo {volume} {65}},\ \bibinfo
  {pages} {165122} (\bibinfo {year} {2002})}\BibitemShut {NoStop}%
\bibitem [{\citenamefont {Moreno}\ \emph {et~al.}(2011)\citenamefont {Moreno},
  \citenamefont {Muramatsu},\ and\ \citenamefont {Manmana}}]{Moreno2011}%
  \BibitemOpen
  \bibfield  {author} {\bibinfo {author} {\bibfnamefont {A.}~\bibnamefont
  {Moreno}}, \bibinfo {author} {\bibfnamefont {A.}~\bibnamefont {Muramatsu}}, \
  and\ \bibinfo {author} {\bibfnamefont {S.~R.}\ \bibnamefont {Manmana}},\
  }\href {\doibase 10.1103/PhysRevB.83.205113} {\bibfield  {journal} {\bibinfo
  {journal} {Phys. Rev. B}\ }\textbf {\bibinfo {volume} {83}},\ \bibinfo
  {pages} {205113} (\bibinfo {year} {2011})}\BibitemShut {NoStop}%
\bibitem [{\citenamefont {GraceQuantum.org}()}]{GraceQ}%
  \BibitemOpen
  \bibfield  {author} {\bibinfo {author} {\bibnamefont {GraceQuantum.org}},\
  }\href {https://mps2.gracequantum.org} {\ }\bibinfo {note} {GraceQ/MPS2: A
  high-performance matrix product state algorithms library based on
  GraceQ/tensor.}\BibitemShut {Stop}%
\end{thebibliography}%

\newpage

\appendix

\setcounter{equation}{0}
\setcounter{figure}{0}
\setcounter{table}{0}
\setcounter{page}{1}
\makeatletter
\renewcommand{\theequation}{S\arabic{equation}}
\renewcommand{\thefigure}{S\arabic{figure}}

\section{Supplemental Material}

\section{More results for lightly doped gapless spin liquid}
\begin{figure}[htb!]
\centering
    \includegraphics[width=1.03\linewidth]{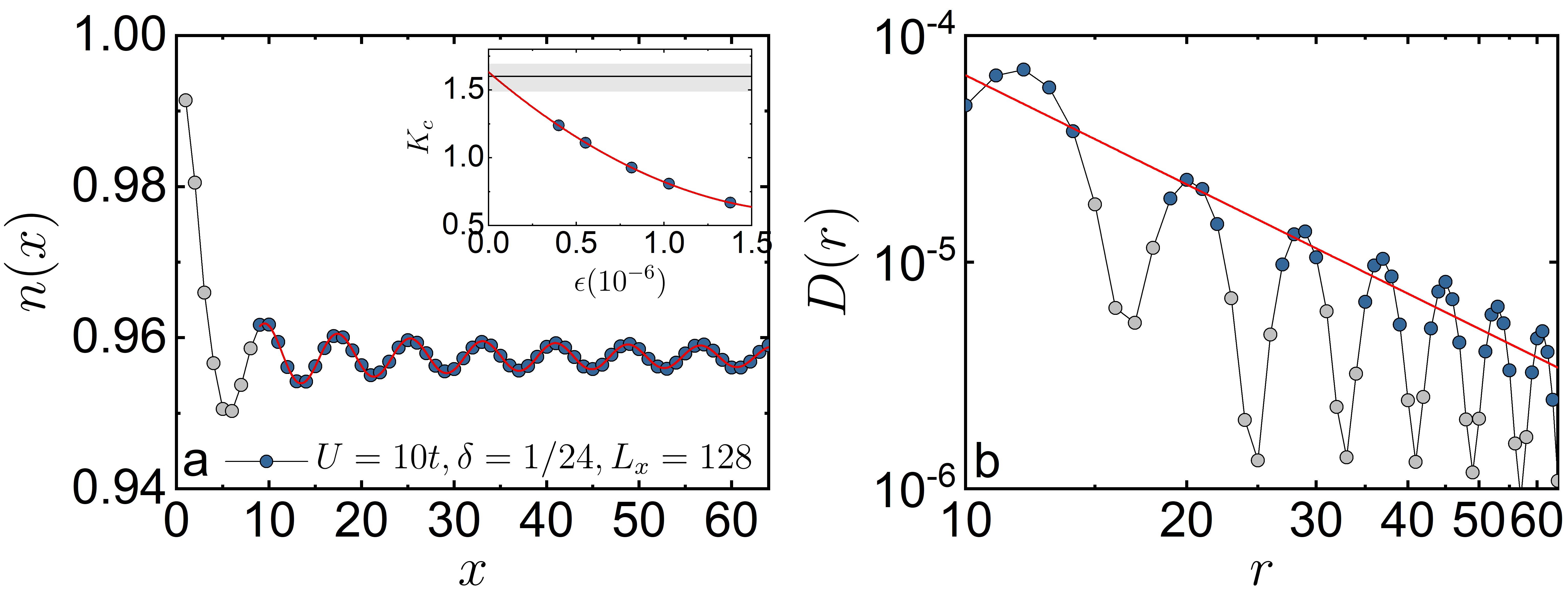}
\caption{(Color online) (a) Charge density profile $n(x)$ for $U=10t$ at the doping level $\delta=1/24$. Inset: Luttinger exponent $K_c$ extracted using Friedel oscillation fitting as a function of truncation error $\epsilon$. (b) density-density correlation $D(r)$ and its algebraic fit $D(r)\sim r^{-K_c}$ labelled by the red line. The $K_c$ extracted from (b) has been shown as a black line with error bar in the inset of (a).}\label{Figs:doped_density}
\end{figure}

We provide more results here for the charge density-density correlation $D(r)$ measured from $U=10t$ at $\delta=1/24$ in the lightly doped gapless spin liquid phase. Alternatively, the Luttinger exponent $K_c$, which is extracted using Eq.\ref{Eq:Friedel}, can also be extracted from the density-density correlation $D(r)$, defined as
\begin{equation}\label{densitycor}
  D(r)=\frac{1}{L_y}\sum_{y=1}^{L_y}|\langle(n_{(x_0,y)}-\rho)(n_{(x_0+r,y)}-\rho) \rangle|,
\end{equation}
where $\rho=1-\delta$ is the electron density. For the system with quasi-long-range CDW order, $D(r)$ should also decay with a power-law $D(r)\sim r^{-K_c}$ where the exponent $K_c$ is identical within the error bar to the one extracted from Eq.\ref{Eq:Friedel} in the thermodynamic limit. Fig.\ref{Figs:doped_density} shows the extracted $K_c$ from both methods $L_x=128$ cylinder. In Fig.\ref{Figs:doped_density}a, the extracted $K_c(\epsilon)$ from the Friedel oscillation as a function of the finite truncation error is $K_c=1.6(1)$. For comparison, Fig.\ref{Figs:doped_density}b shows the density-density correlation $D(r)$ of the same cylinder, where the extracted exponent is also $K_c=1.6(1)$, which is as expected consistent with that obtained by the Friedel oscillation the within the error bar.

\begin{figure}[htb!]
\centering
    \includegraphics[width=0.9\linewidth]{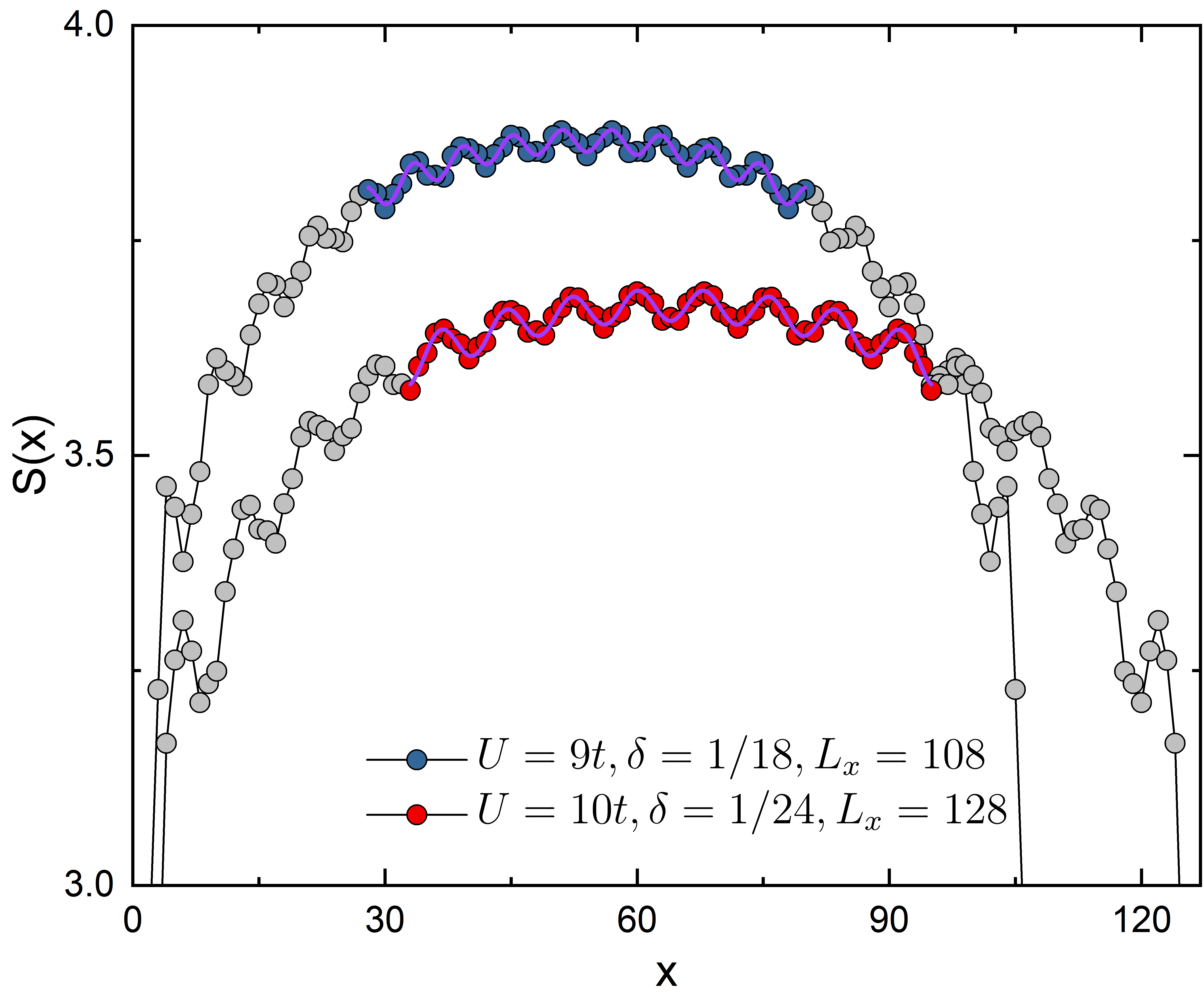}
\caption{(Color online) von Neumann entanglement entropy $S(x)$ on cylinders. Solid lines are fitted from the Eq.\ref{Eq:central_charge} using the entanglement entropy on the central part of the cylinders. Data points in gray are removed to minimize the boundary effect.}\label{Figs:EntropyOscillation}
\end{figure}

We can estimate the central charge $c$ for (quasi-) 1D system of length $L_x$ using a more precise formula\cite{Calabrese2004,Fagotti2011}, 
\begin{eqnarray}
S(x)&=&\frac{c}{6} \ln \left[\frac{4(L_x+1)}{\pi} \sin \frac{\pi(2x+1)}{2(L_x+1)}
\right] \nonumber\\
&+& \frac{A\sin[q(2x+1)]}{\frac{4(L_x+1)}{\pi} \sin \frac{\pi(2x+1)}{2(L_x+1)}
}+ \tilde{c}. \label{Eq:central_charge}
\end{eqnarray}
Here $A$, $\tilde{c}$ and $q$ are model dependent constants. As shown in Fig.\ref{Figs:EntropyOscillation}, the estimated central charge is $c=1.3(1)$ for $U=9t$ at $\delta=1/18$, and $1.2(1)$ for $U=10t$ at $\delta=1/24$, respectively. This is fairly close to $c=1$ which suggests that there is probably single gapless charge mode.


\section{Lightly doped Dimer phase }
\begin{figure}[htb!]
\centering
    \includegraphics[width=1\linewidth]{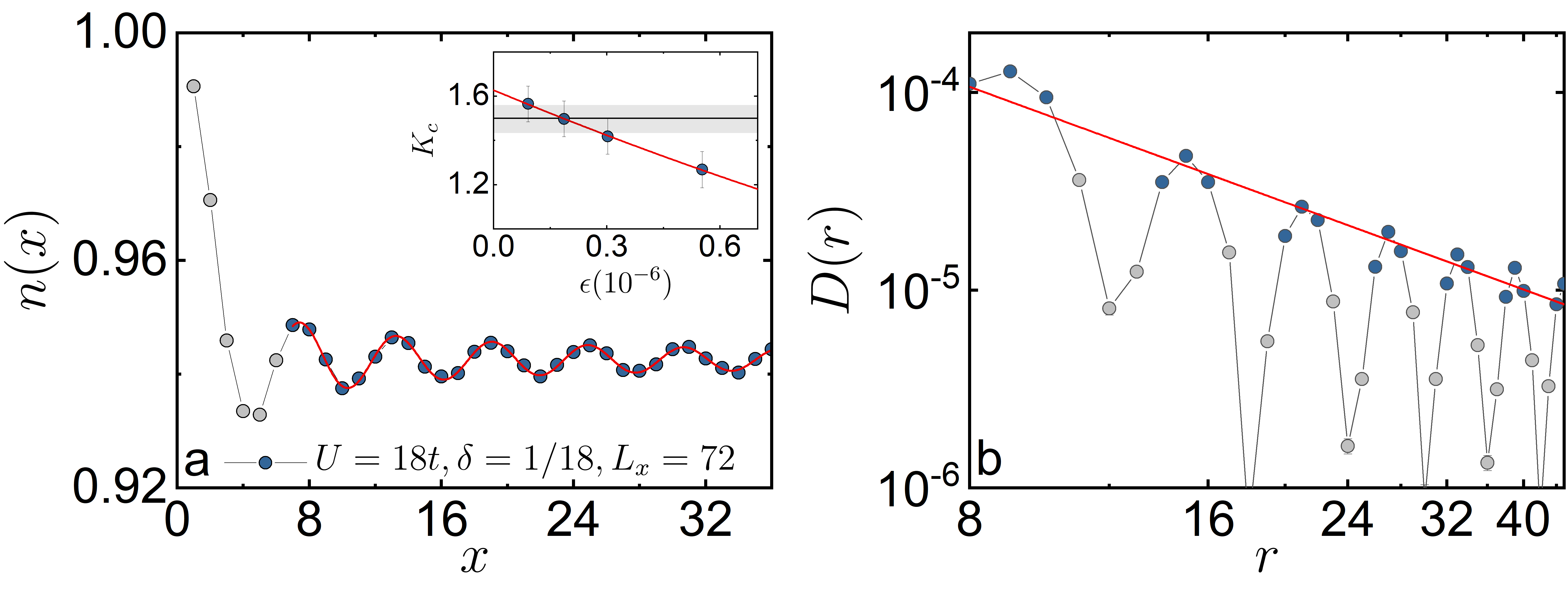}
\caption{(Color online) (a) Charge density profile $n(x)$ for $U=18t$ at the doping level $\delta=1/18$. Inset: Luttinger exponent $K_c$ extracted using Friedel oscillation fitting as a function of truncation error $\epsilon$. (b) density-density correlation $D(r)$ and its algebraic fit $D(r)\sim r^{-K_c}$ labelled by the red line. The $K_c$ extracted from (b) has been shown as a black line with error bar in the inset of (a).}\label{Figs:U18HD18CDW}
\end{figure}

\begin{figure}[htb!]
\centering
    \includegraphics[width=1\linewidth]{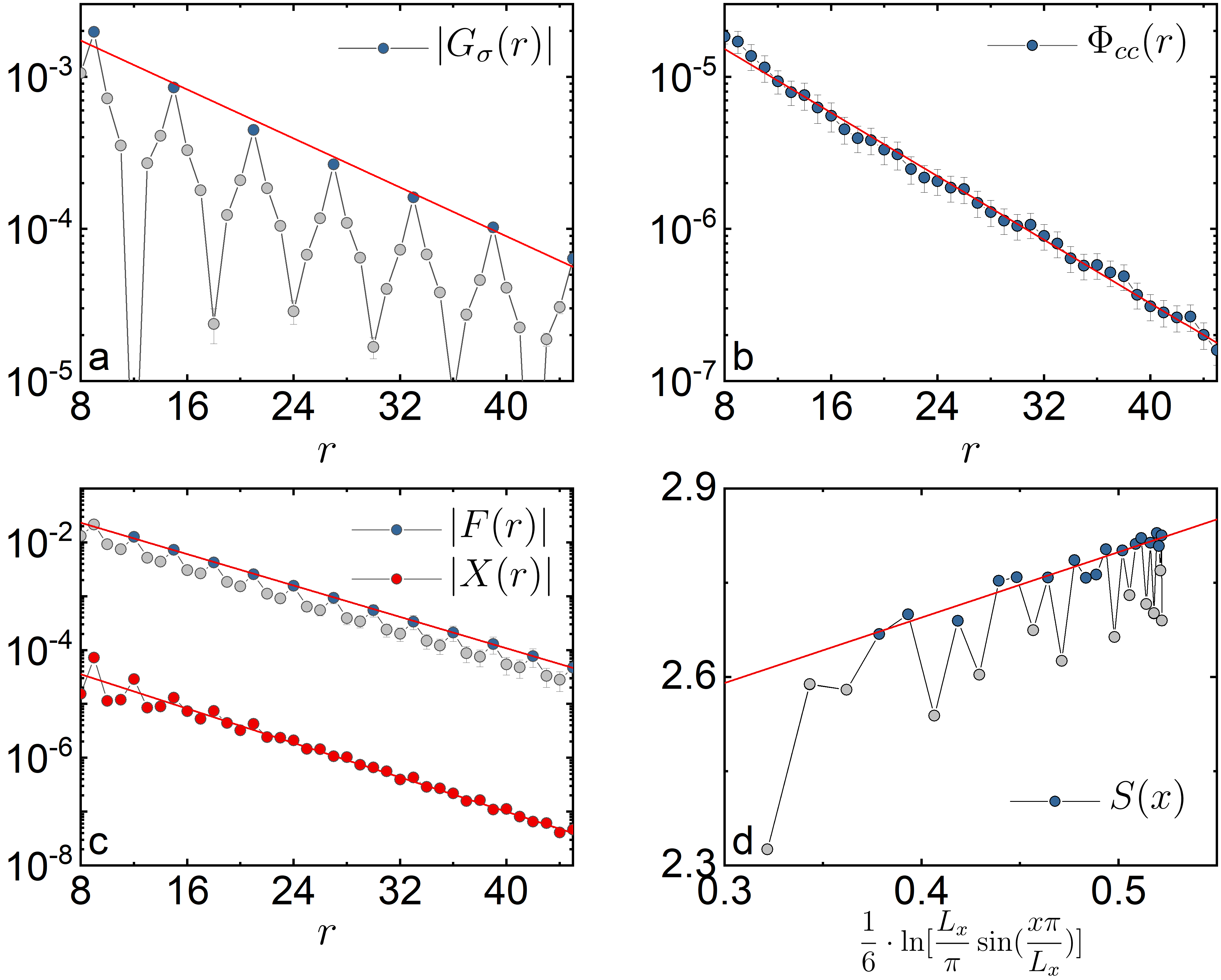}
\caption{(Color online) Correlation functions for $U=18t$ at $\delta=1/18$ doping. (a) The single particle correlation function in log-linear plot. The red line denotes the fitting curve $|G_{\sigma}(x)|\sim e^{-r/\xi_{G}}$. (b)The pair-field correlation function $\Phi_{cc}(r)$ and the exponential fitting curves. (c) The spin-spin and scalar chiral-chiral correlation functions and the exponential fitting curves. (d) The entanglement entropy $S(x)$ and the fitting curve.}\label{Figs:U18HD18CorlFunc}
\end{figure}

To characterize the ground-state properties of lightly doped dimer phase, we consider $U=18t$ as an example which is deep inside the dimer phase and consider a typical doping concentration $\delta=1/18$. As shown in Fig.\ref{Figs:U18HD18CDW}a, the charge density distribution $n(x)$ has a well-defined ordering wavevector $K=6\pi\delta$ in the $\mathbf{e}_1$ direction and can be fitted by the Freidel oscillation (Eq.\ref{Eq:Friedel}) with $K_c=1.5(1)$ in the $\epsilon=0$ limit. Alternatively, $K_c$ can also be obtained from the algebraic fitting of $D(r)\sim r^{-K_c}$, as shown in Fig.\ref{Figs:U18HD18CDW}b, with the extracted exponents $K_c=1.5(1)$, which is also consistent with that obtained from the Freidel oscillation.

We have also calculated various other correlation functions as shown in Fig.\ref{Figs:U18HD18CorlFunc} and find that they all decay exponentially at long distances. These include the single-particle correlation $G_\sigma(r)\sim e^{-r/\xi_G}$, the superconducting correlation $\Phi_{\alpha\beta}(r)\sim e^{-r/\xi_{sc}}$, the spin-spin correlation $F(r)\sim e^{-r/\xi_s}$, and the scalar chiral-chiral correlation $X(r)\sim e^{-r/\xi_\chi}$.
The corresponding correlation lengths are summarized in the main text in Table.\ref{Table:IntMidQSExponent}.

Aside from the correlation functions, we have also calculated the central charge $c$. For the critical (quasi-) 1D system, the von Neumann entanglement entropy of the subsystem follows the formula $S(x) = \frac{c}{6}\ln[\frac{L_x}{\pi}\sin(\frac{\pi x}{L_x})] + {\rm const}$, with the central charge $c$ equals to $1$. As shown in Fig.\ref{Figs:U18HD18CorlFunc}d, the central charges extracted is $c=1.04(7)$. This is consistent with single gapless charge mode.

\section{Compare results with complex DMRG code}
\begin{figure}[tb!]
\centering
    \includegraphics[width=1\linewidth]{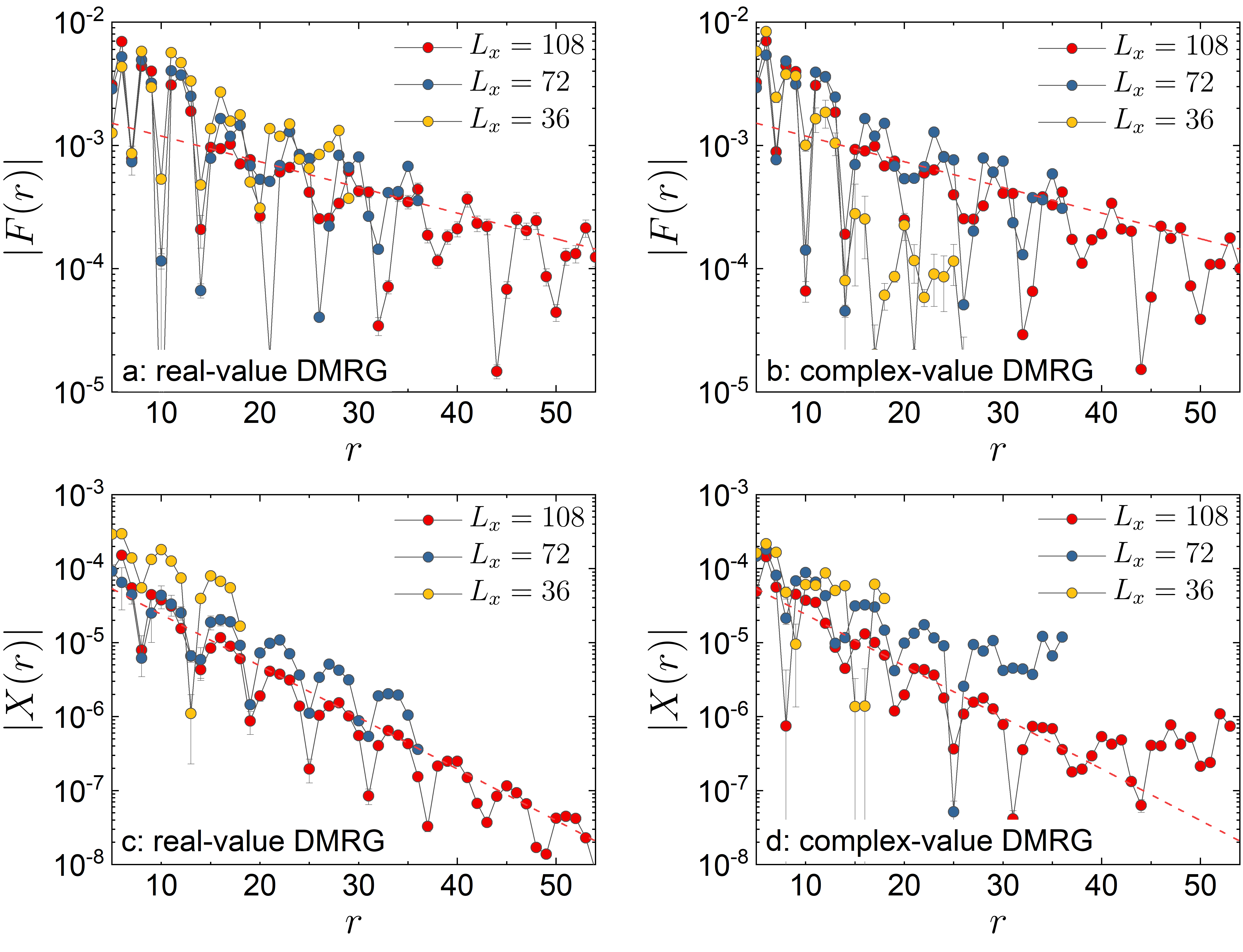}
\caption{(Color online) Spin-spin and scalar chiral-chiral correlation correlation calculated from real-value and complex-value DMRG code for $U=9t$ at $\delta=1/18$ doping. The dashed lines are showing the fitting functions of the real-value DMRG on $L_x=108$ cylinder. }\label{Figs:U9HD18CorlFuncCplx}
\end{figure}

Previous study\cite{Zhu2020} suggests lightly doping the intermediate phase could lead to a chiral metal phase, which spontaneously breaks the time-reversal symmetry. As a result, an important numerical check which needs to be done is whether real-value and complex-value DMRG simulations provide the qualitatively same results. In this section, we provide direct evidences that both DMRG simulations indeed gives us the similar results, both qualitatively and quantitatively. As examples, we have calculated both the spin-spin $F(r)$ and scalar chiral-chiral correlations $X(r)$ on $L_x=36\sim 108$ systems by keeping up to $m=25000$ number of states in the real-value DMRG simulation and up to $m=16000$ number of states in the complex-value DMRG calculation. As shown in Fig.\ref{Figs:U9HD18CorlFuncCplx}, while the results obtained from complex-value DMRG simulations suffer from a significantly larger boundary effects for relatively small systems (such as $N=36\times3$), they are clearly consistent with the results obtained from real-value DMRG simulations, including both $F(r)$ and $X(r)$. For both DMRG simulations, it is clear that $F(r)$ and $X(r)$ decay exponentially and can be well-fitted by an exponential function as $F(r)\sim e^{-r/\xi_s}$ and $X(r)\sim e^{-r/\xi_\chi}$. Moreover, we have checked and calculated other correlations, including single particle and SC correlations, where both real- and complex-value DMRG simulations give us the same results.

\end{document}